\documentclass[numberedappendix]{emulateapj}
\usepackage{graphicx}
\usepackage{color}

\begin{document}


\newcommand{\lsim}{\lower0.6ex\vbox{\hbox{$ \buildrel{\textstyle <}\over{\sim}\ $}}}
\newcommand{\gsim}{\lower0.6ex\vbox{\hbox{$ \buildrel{\textstyle >}\over{\sim}\ $}}}
\newcommand{\scb}{\textcolor{blue}}
\newcommand{\scr}{\textcolor{red}}


\newcommand{\hvol}{h^{3}{\mathrm{Mpc}}^{-3}}
\newcommand{\hmpc}{h^{-1}\mathrm{Mpc}}
\newcommand{\hkpc}{h^{-1}\mathrm{kpc}}
\newcommand{\hpc}{h^{-1}\mathrm{pc}}
\newcommand{\hMsun}{\ h^{-1}\mathrm{M}_{\odot}}
\newcommand{\hMpc}{\ h^{-1}\mathrm{Mpc}}
\newcommand{\Msun}{M_{\odot}}
\newcommand{\kms}{{\,{\mathrm{km}}\,{\mathrm{s}}^{-1}}}
\newcommand{\kpc}{{\,{\mathrm{kpc}}}}
\newcommand{\Gyr}{{\,{\mathrm{Gyr}}}}

\newcommand{\rhomean}{\rho_{\mathrm{M}}}

\newcommand{\mpt}{m_{\mathrm{p}}}
\newcommand{\rfind}{r_{\mathrm{f}}}


\newcommand{\Mr}{M_r^h}
\newcommand{\Mvir}{M_{\mathrm{vir}}}
\newcommand{\Rvir}{R_{\mathrm{vir}}}
\newcommand{\Dvir}{\Delta_{\mathrm{vir}}}
\newcommand{\Vmax}{V_{\mathrm{max}}}

\newcommand{\Vmaxa}{V_{\mathrm{max, acc}}}
\newcommand{\Vmaxn}{V_{\mathrm{max, now}}}

\newcommand{\vmax}{V_{\mathrm{max}}}
\newcommand{\vmaxa}{V_{\mathrm{max, acc}}}
\newcommand{\vmaxn}{V_{\mathrm{max, now}}}

\newcommand{\Vhost}{V_{\mathrm{host}}}
\newcommand{\Vsat}{V_{\mathrm{sat}}}

\newcommand{\Nsat}{N_{\mathrm{sat}}}
\newcommand{\Ngal}{N_{\mathrm{gal}}}

\newcommand{\LCDM}{\Lambda CDM}

\shortauthors{Felipe Mar\'in}
\shorttitle{SDSS LRG 3PCF bias}

\title{The Large-Scale SDSS LRG 3-point correlation function}

\author{Felipe Mar\'in\altaffilmark{1,2}}
\altaffiltext{1}{Department of Astronomy \& Astrophysics and Kavli Institute for Cosmological Physics, The
 University of Chicago, Chicago, IL 60637 USA} 
\altaffiltext{2}{Centre for Astrophysics \& Supercomputing, Swinburne University
of Technology, PO Box 218, Hawthorn, VIC 3122, Australia; fmarin@astro.swin.edu.au}

\journalinfo{Submitted to The Astrophysical Journal}

\begin{abstract}

We present new measurements of the redshift-space three-point 
correlation function (3PCF) of Luminous Red Galaxies (LRGs)
from the Sloan Digital Sky Survey (SDSS). Using the largest dataset to date, the Data
Release 7 (DR7) LRGs, and 
an improved
binning scheme compared to previous measurements, we
measure the LRG 3PCF on large scales up to $\sim 90 ~ \hmpc$, 
from the mildly non-linear to quasi-linear regimes. Comparing the LRG correlations to 
the dark matter two- and three-point correlation functions, obtained from N-body simulations
we infer linear and non-linear bias parameters. As expected, LRGs are highly biased
tracers of large scale structure, with a linear bias $b_1 \sim 2$; the LRGs also have a large
positive non-linear bias parameter, in agreement with predictions of galaxy population models.
The use of the 3PCF to estimate biasing helps to also make estimates of the 
cosmological parameter $\sigma_8$, as well as to infer best-fit parameters
 of the Halo Occupation Distribution 
parameters for LRGs.
We also use a large suite of public mock catalogs to characterize 
the error covariance matrix for the 3PCF and compare the variance 
among simulation results with jackknife error estimates.  

\end{abstract}


\section{Introduction}

The large-scale structure traced by galaxies is shaped both by 
cosmic expansion history, which determines the gravitational evolution 
of density perturbations, and 
by the physics of galaxy formation. Comparison 
of galaxy clustering with the predictions of structure formation models 
therefore constrains both cosmological and galaxy formation 
parameters.  For a given clustering observable, e.g., the 
galaxy two-point correlation function (2PCF), $\xi_{gg}(r)$, there are typically 
significant degeneracies between the inferred 
cosmological and galaxy formation 
parameters. Measurement of multiple observables with different relative 
dependencies on cosmology and galaxy formation can help break such 
degeneracies \citep{abazajian_etal:05,zheng_weinberg:05}. The 
galaxy three-point 
correlation function (3PCF), the next level up in the hierarchy of $N$-point correlation 
functions, encodes information that is complementary to that contained 
in the 2PCF and is therefore a useful second observable for constraining 
cosmology and galaxy formation \citep{sefusatti_scoccimarro:05,sefusatti_etal:06}.    

The 3PCF has been measured since the advent of the first
angular catalogs \citep{peebles_groth:75}, and more recently in the last generation of 
spectroscopic surveys such as 2dFGRS \citep{jing_borner:04, gaztanaga_etal:05},  and SDSS 
\citep{kayo_etal:04,nichol_etal:06,kulkarni_etal:07,mcbride_et_al:10b}. The goals of
these efforts have been  mostly to test predictions from theories of growth of structure and 
cosmological simulations, as well as to measure the biasing of the galaxies with respect to the dark
matter 3PCF. With the large volume of current surveys, we are able to improve the
signal-to-noise of the measurements, and obtain more information from the 3PCF to help constrain
cosmological and galaxy formation models.

In this paper, we measure the 3PCF for a particular class of galaxies, the 
Luminous Red Galaxies (LRGs) \citep{eisenstein_etal:01} observed spectroscopically by the 
Sloan Digital Sky Survey (SDSS) \citep{york_etal:00}, and use the results to 
constrain galaxy bias and cosmological parameters. 
LRGs offer several advantages 
for this measurement. At the bright end of the 
galaxy luminosity function, LRGs can be targeted out to relatively large 
distances, $z \sim 0.5$ (whereas the average redshift is $z\sim0.3$), 
in the SDSS spectroscopic survey. Their clustering can therefore 
be measured over scales large compared to those probed by $L_*$ galaxies in the 
same survey \citep{eisenstein_etal:05,tegmark_etal:06,percival_etal:07,gaztanaga_etal:08b},
which are on average at $z\sim0.1$. 
In addition, LRGs are more strongly clustered than less-luminous 
galaxies, making their correlations intrinsically easier to measure on large 
scales, where shot-noise is subdominant.

The physics of galaxy formation is imprinted in the $N$-point galaxy 
correlation functions via the bias, the relation between  
the spatial distributions of galaxies and dark matter. Although the
Halo Occupation Distribution  model (HOD, see for instance
 \citealt{jing_etal:98, seljak:00,scoccimarro_etal:01, berlind_weinberg:02,cooray_sheth:02, zheng_etal:05}) 
has become a popular 
framework for parametrizing the bias (especially on small scales), here we focus on using the 
2PCF and 3PCF LRG measurements on large scales to constrain a simpler, 
non-linear, local bias model that appears to adequately capture the
features displayed by the clustering data. 
This approach allows us to break the degeneracy that exists between
the linear bias parameter and the r.m.s of density fluctuations at scale of 8 $\hmpc$, $\sigma_8$,
 when extracting
information from the 2PCF, as shown, using different statistics, in \cite{pan_szapudi:05} and  \cite{ross_etal:08}.
Although a complete and detailed HOD analysis using fits to the 2-point and 3-point functions will be
presented elsewhere \citep{sabiu_etal:10}, we explore the idea of  using  the LRG bias parameters in constraining
HOD models, as suggested by \cite{sefusatti_scoccimarro:05}.

We note that \cite{gaztanaga_etal:08a} have recently measured 
the SDSS LRG 3PCF on very large scales, focusing on detection of the baryon 
acoustic oscillation signature. \cite{reid_spergel:08} used a counts-in-cylinders 
statistic to constrain the LRG HOD.
In addition, \cite{ross_etal:08} measured higher-order 
angular clustering of a sample of photometrically selected LRGs from the SDSS and used it 
to constrain non-linear bias parameters and get constraints on $\sigma_8$. 

Our goal here is to present an up-to-date measurement of the 3PCF, show another way to use
the 3PCF information  
on large scales, in this case to calculate the bias, and offer complementary constraints
on $\sigma_8$, and show that we can use the 3PCF as a complement to find best-fit HOD parameters.

The outline of the paper is as follows.
In \S \ref{sec:methods} we 
review the 3PCF and describe our binning scheme for the 3PCF measurement; we also 
describe the SDSS LRG sample and the simulations used to create mock catalogs.
In \S \ref{sec:lrg3pcf} we present our measurements
of the 3PCF in the LRG sample.
In \S \ref{sec:errors} we discuss the estimation of 3PCF errors, 
which are critical in constraining parameters. 
In \S \ref{sec:bias} we constrain the non-linear bias parameters of the LRGs and 
explore the possibility
of constraining the matter power spectrum amplitude $\sigma_8$, as well as relevant HOD parameters.
We summarize and conclude 
in \S  \ref{sec:conclusions}.

\section{Methods, Data \& Simulations}
\label{sec:methods}

\subsection{The 3-point correlation function}
\label{sec:3pcf}

\begin{figure}
\plotone{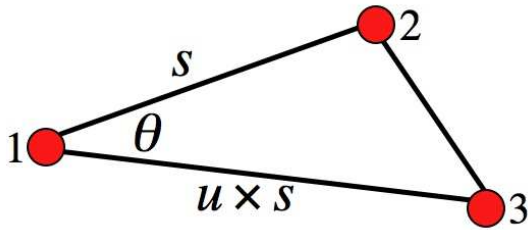}
\caption{Parametrization of triangles for measuring the three-point correlation 
function.}
\label{fig:triangle}
\end{figure}

The 3PCF describes the probability of finding three
objects in a particular triangle configuration, compared to that of a random
sample. The joint probability of finding three objects in three infinitesmal volumes
$dV_1$, $dV_2$, and $dV_3$ is given by \citep{peebles:80}
\begin{eqnarray}\label{eq:P3}
P & = & [1+\xi(r_{12})+\xi(r_{23})+\xi(r_{31})+\zeta(r_{12},r_{23},r_{31})] \times \nonumber \\
&& \bar{n}^3 dV_1 dV_2 dV_3 ~,
\end{eqnarray}

where $\bar{n}$ is the mean density of objects, $\xi$ is the
2PCF, and $\zeta$ is the (connected) 3PCF:
\begin{eqnarray}
\xi(r_{12}) &=& \langle\delta(r_1)\delta(r_2)\rangle \\
\zeta(r_{12},r_{23},r_{31})&=&\langle\delta(r_1)\delta(r_2)\delta(r_3)\rangle;
\end{eqnarray}
where $\delta$ is the fractional overdensity of objects or the continuous field studied.
The triangle sides $r_{ij}$  
are the distances between objects $i$ and $j$ in the triplet. 
Since the 3PCF depends
upon the configuration of the three sides, it is sensitive to the
shapes of spatial structures   
\citep{sefusatti_scoccimarro:05,gaztanaga_scoccimarro:05,marin_etal:08}.
Since the ratio $\zeta/\xi^2$  is both predicted and found to be close to
unity over a large range of length scales even though $\xi$ and $\zeta$ each    
vary by orders of magnitude \citep{peebles:80}, it is convenient to 
define the reduced 3PCF, 
\begin{equation}
Q(s,u,\theta) \equiv \nonumber \frac{\zeta(s,u,\theta)}{\xi(r_{12})\xi(r_{23})+ \xi(r_{23})\xi(r_{31})+\xi(r_{31})\xi(r_{12})}~.
\end{equation}
Here, $s\equiv r_{12}$ sets the scale size of the triangle, and the shape
parameters are given by the ratio of two sides of the triangle, 
$u \equiv r_{23}/r_{12}$, and the angle between those two sides,
$\theta=\cos^{-1}(\hat{r}_{12}\cdot \hat{r}_{23})$, 
where $\hat{r}_{12}$, $\hat{r}_{23}$ are unit vectors in the directions of 
those sides (see Figure \ref{fig:triangle}).
By measuring the shape- and scale-dependence of $Q(s,u,\theta)$ for galaxies and 
comparing with that predicted for dark matter, we can constrain the galaxy bias.

We calculate
the 2PCF using the estimator of \cite{landy_szalay:92}, 
\begin{equation}
\xi = \frac{DD-2DR+RR}{RR}.
\end{equation}
Here, $DD$ is the number of data pairs normalized by $N_D\times
N_{D}/2$, $DR$ is the number of pairs using data and random catalogs
normalized by $N_DN_R$, and $RR$ is the number of 
random data pairs normalized by
$N_R\times N_R/2$, where $N_D$ and $N_R$ are the number of points in the data and
in the random catalog, respectively.  The 3PCF is calculated using the
\cite{szapudi_szalay:98} estimator:
\begin{equation}
\zeta=\frac{DDD-3DDR+3DRR-RRR}{RRR},
\end{equation}
where $DDD$, the number of data triplets, is normalized by $N_D^3/6$,
and $RRR$, the random data triplets, is normalized by $N_R^3/6$. $DDR$ is
normalized by $N_D^2N_R/2$, and $DRR$ by $N_DN_R^2$.
These estimators optimally correct 
for edge effects by incorporating pair and triplet counts from a random catalog.  
To compute the correlation functions, we use the {\it NPT} software developed
in collaboration with the Auton Lab at Carnegie Mellon
University. {\it NPT} is a fast implementation of $N$-point correlation 
function estimation using
multi-resolution kd-trees to compute the number of pairs and
triplets in a dataset. For more details and information on the
algorithm, see \cite{moore_etal:01}, \cite{gray_etal:04}, and
\citet{nichol_etal:06}.

\subsection{Measuring the 3PCF: Binning}
\label{sec:binning}

\begin{figure}
\plotone{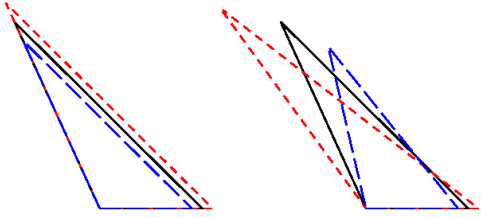}
\caption{Comparison of two binning schemes for the 3PCF measurement. {\it Left:} bins 
in $r_{ij}$ used in this paper, with $\Delta r_{ij}=0.1 r_{ij}$; 
{\it Right:} bins in $s$, $u$, $\theta$, with $\Delta s=0.1s$, 
$\Delta u=0.1$, $\Delta \theta = \pi/100$. 
Black (solid) triangles show the central configuration for the bin, with 
$u=2$ and $\theta = 100^o$. 
Blue (long dashed) triangles are the smallest triangles  
in that bin. Red (short dashed) triangles are the largest in the bin. 
For the $r_{ij}$ bins, the largest and smallest triangles in the bin are 
similar to the central triangle.}
\label{fig:shapes}
\end{figure}

\begin{figure}
\plotone{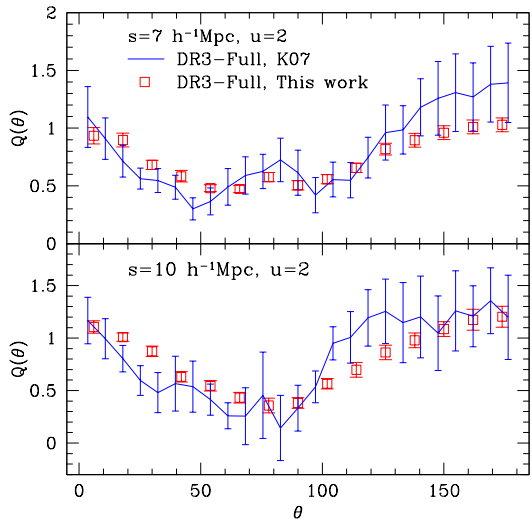}
\caption{Comparison of two binning schemes the DR3 LRG 3PCF measurement.
DR3 Sample LRG reduced 3PCF in for $s=7$ $\hmpc$ (top)  and $s=10$ $\hmpc$ (bottom)
with similar resolution but different binning schemes. Blue solid line represents 3PCF from the
scheme used in \cite[][K07 in the plot]{kulkarni_etal:07} and open squares used the binning representation used
in this paper. Diagonal errors obtained from jack-knife resampling method.}
\label{fig:kulcomp}
\end{figure}

Since measurements of two- and three-point correlations involve 
counts of pairs and triplets of objects, the question arises of how to choose the 
bins of pair and triplet separation \citep{gaztanaga_scoccimarro:05, nichol_etal:06,
kulkarni_etal:07, marin_etal:08}. Small bins enable measurements with 
fine resolution in separation and the possible identification of 
features in the correlation functions but at the cost of large shot noise due 
to the small numbers of pairs and triplets per bin. Large bins reduce the shot 
noise but tend to wash out features of the 3PCF. The optimal bin size is therefore a 
compromise and depends on the relative importance of signal-to-noise vs. 
resolution for the question at hand. 

For the 3PCF, there is also the 
question of which separation parameters to bin in, that is, which triangle 
configurations to include in each bin.  \cite{nichol_etal:06} and 
\cite{kulkarni_etal:07} measured the 3PCF in bins of $s$, $u$, and $\theta$, that is, 
they counted triplets of galaxies that satisfy $s \in s_0 \pm \Delta s$, 
$u \in u_0 \pm \Delta u$, $\theta \in \theta_0 \pm \Delta \theta$ for the 
bin centered on $u_0$, $s_0$, $\theta_0$.  In their measurement of the LRG 3PCF using
the DR3 sample, \cite{kulkarni_etal:07} used very small resolution in $\Delta s=\pm0.1$ $Mpc$ (1\% at 
10 $\hmpc$) and $\Delta \theta=\pm \pi/100$, and an intermediate value for $\Delta u=\pm0.1$. With such small 
resolution the result is that $Q(\theta)$ is measured with low signal-to-noise (since there are few pairs and triplets
in each bin) and since $\Delta u$ is not small (for $u=2$, $s=10$ the bin size of the second side of the triangle is 12\% of the side),
there are triplets counted in two or more configurations, or 'bins'. With this binning scheme is not possible to measure the LRG 3PCF on larger scales.
But just increasing bin size in this way would lead to have very different triangles accepted in the same bin, as shown 
in the right side of  Fig. \ref{fig:shapes}.

In this work, we use the parameters $s$, $u$ and $\theta$ to determine the centers of the triplet bins, 
but we construct the bins themselves 
directly in each of the pair separations, with  $\Delta r_{ij} \propto r_{ij}$, i.e., 
we use bins of constant $\Delta \log r_{ij}$  \citep{marin_etal:08,mcbride_etal:10,mcbride_et_al:10b}.  
As Fig. \ref{fig:shapes} shows, this binning scheme groups together triplet configurations 
that are more similar in shape (both mathematically and colloquially) and size than do bins 
in $s$, $u$, and $\theta$.  As a result, we can achieve either higher resolution in 
triangle shape and size or higher signal-to-noise for fixed resolution. Also to obtain a better signal in the 3PCF, 
we increase somewhat the size of the bins, which leads to a bigger correlation between the points, and 
therefore we do not need to measure the 3PCF for a large number of angles $\theta$ as \cite{kulkarni_etal:07} did.

 We show in Figure
\ref{fig:kulcomp} the results of using the different binning schemes: using the DR3 LRG sample
(see \S \ref{sec:sample} for a description of the sample), we compare the resultant 3PCF using
the binning scheme used in \cite[][in green solid lines]{kulkarni_etal:07}, 
with the scheme used in this paper, for $\Delta r_{ij}=\pm0.1r_{ij}$ (in open squares); 
it is clear that in our binning scheme $Q(\theta)$ fluctuates
much less than when using the 'old' scheme, it reproduces better the theoretical predictions (note
 that there is no plateau in the
new scheme that happens in the $\theta >100$ in the $s=10$ $\hmpc$ configurations), and has smaller
errors (calculated, as in \cite{kulkarni_etal:07}, using jack-knife resampling, which will be discussed in \S \ref{sec:errors}).

\begin{figure*}
\plotone{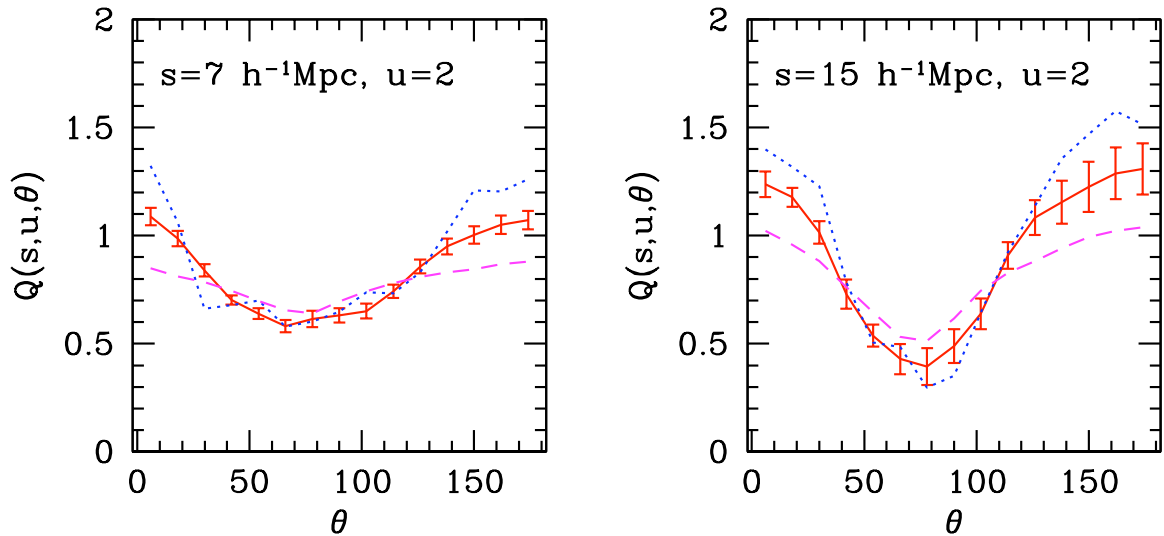}
\caption{Effects of binning in the 3PCF measurement for the SDSS DR7-Dim
  LRG sample, for configurations with $u=2$. Left:
  measurements with $s=7 \hmpc$. Right: $s=15$ $\hmpc$. 
The blue dotted, solid red, and dashed magenta curves represent results
with $\Delta r_{ij}/r_{ij}=\pm 0.05$, 0.1, and 0.2,
respectively. }
\label{fig:bincomp}
\end{figure*}

We take advantage of this  binning scheme to measure the 3PCF 
over a large range of triangle sizes and shapes: 
we study central configurations with $s$ ranging from 7 (which allows us to compare this measurements to 
previous works) to 30 $\hmpc$ (on the quasi-linear regime, in order to estimate bias parameters),
 with $u=2$ and 15 equally spaced values of $\theta$.  For other studies (outside the scope of this paper) such
 as finding best-fit galaxy population (HOD) parameters from fits to the 3PCF, it is necessary to carry out measurements
 on smaller scales.
The results are limited by shot noise on 
small scales (due to the small spatial density of the LRG field)
 and by finite volume on large scales.

In Figure \ref{fig:bincomp} we illustrate the sensitivity of the
reduced 3PCF measurement to the choice of binning resolution. We 
show measurements with $s=7$ and $15$ $\hmpc$ for the SDSS 
DR7-Dim LRG sample (see \S \ref{sec:sample}).
The blue dotted, solid red, and dashed magenta curves represent results
with $\Delta r_{ij}/r_{ij}=\pm 0.05$, 0.1, and 0.2, respectively. 
We can see significant bin-dependent differences in the reduced 3PCF amplitude, 
particularly for elongated triangle configurations ($\theta$ close
to $0^\circ$ or 180 $^\circ$). The differences appear to increase  for larger
scales.  Proceeding from smaller to larger bins, the
configuration-dependence of $Q$ is smoothed out, and the measurements
for different values of  $\theta$ become more strongly correlated. On
the other hand, the signal-to-noise for each measurement increases
with bin width. We choose
our binning to balance the need for a good 3PCF signal with the
ability to discern shape-dependent features
of the 3PCF. After experimenting with simulations, 
 we settled on a 14\%, i.e. $\Delta r = \pm 0.07r$
binning resolution, for all LRG samples we studied (see \S \ref{sec:sample}).

\begin{deluxetable*}{l c c c c c}
\tablecaption{SDSS LRG Samples used}
\tablehead{
\colhead{Sample} & \colhead{area} &
\colhead{redshift} & 
\colhead{Luminosity} & \colhead{Number of galaxies}&\colhead{average density}}
DR7--Full \footnote{from \cite{kazin_etal:09}}  & 7,908 deg$^{-2}$& $0.16<z<0.47$&$-21.2>M_g>-23.2$&105,831&$6.70\times10^{-5}$ $(\hmpc)^3$\\
DR7--Dim $^a$ & 7,189 deg$^{-2}$& $0.16<z<0.36$&$-21.2>M_g>-23.2$&61,899&$9.40\times10^{-5}$ $(\hmpc)^3$\\
DR7--Bright $^a$& 7,189 deg$^{-2}$& $0.16<z<0.45$&$-21.8>M_g>-23.2$&30,272&$2.54\times10^{-5}$ $(\hmpc)^3$\\
DR3--Full \footnote{from \cite{eisenstein_etal:05}}& 3,816 deg$^{-2}$& $0.16<z<0.55$&$-21.2>M_g>-23.2$&50,987&$6.50\times10^{-5}$ $(\hmpc)^3$
\label{tab:samples}
\end{deluxetable*}


\subsection{The HOD Model}
\label{sec:hodmodel}

The HOD model \citep[eg.]{jing_etal:98, seljak:00,scoccimarro_etal:01, berlind_weinberg:02,cooray_sheth:02, zheng_etal:05}
provides a parametrized prescription for the galaxy spatial 
distribution. The first component is a model for the distribution of dark halos. 
One specifies the halo mass function, $n(M)$, 
the spatial correlations of the halos, and the density profiles for 
halos of mass $M$, all based on analytical models and  fits to N-body simulations
\citep[eg.][]{press_shechter:74, white_rees:78, sheth_etal:01,tinker_etal:08}.
The  second component is the HOD itself, a model specifying how galaxies occupy 
dark halos as a function of halo mass. 

The HOD is a parameterized model of the PDF $P(N|M)$, the probability that 
a halo of mass $M$ contains $N$ galaxies with specified properties. 
The mean occupation function, 
$\langle N(M)\rangle$, is well modeled when one separates the contribution from  central galaxies, $N_{cen}(M)$, 
which is roughly a step function in halo mass (for halos above a certain mass $M_{min}$), 
and the contribution of the satellite 
galaxies, $N_{sat}(M)$, which appears to be well characterized by 
a power law in halo mass \citep{guzik_seljak:02, kravtsov_etal:04,zheng_etal:05} i.e.,
$N_{sat} \sim (M/M_1)^\alpha$, where $M_1$ characterizes the mass of 
halos that host satellite galaxies, and the exponent $\alpha$ describes the high-mass slope 
of the satellite occupation number. 

There is not a unique model for the galaxy HOD; in fact
there are many functional forms of $\langle N(M)\rangle$, as well as the number of 
parameters describing a particular HOD.
Also, apart from the mean HOD, one needs to specify the higher moments of $P(N|M)$.
The distributions of $N_{cen}(M)$ and $N_{sat}(M)$ 
are found to be consistent with nearest-integer (Bernoulli) and Poisson distributions
\citep{kravtsov_etal:04,zheng_etal:05};
and the satellite galaxies are  assumed to be distributed within 
halos according to the NFW halo mass profiles \citep{navarro_etal:97}.
The HOD parameters can be found using different statistics
of  galaxy clustering  \citep[eg.][]{zehavi_etal:03,zehavi_etal:05,kulkarni_etal:07,blake_etal:08,reid_spergel:08, zheng_etal:08,ross_brunner:09}. 

In the case of LRGs, there is consensus  \citep[eg.][and references therein]{kulkarni_etal:07,zheng_etal:08,sabiu_etal:10} that
these galaxies form mostly in high-mass halos with  mass $M \sim$ a few times $10^{13}$ to $10^{14}$ $\hMsun$, while a few percent of them 
(in halos with M $\gsim 10^{14}$ $\hMsun$)
have to be satellites  within halos in order to produce the 2PCF exhibited on smaller scales \citep{masjedi_etal:06}; 
the exact values differ depending on the statistic
used and models of $\langle N(M)\rangle$ that were used. As we will see below in \S \ref{sec:sims}, the HOD model
is a powerful way to create mock galaxy catalogs, and in \S \ref{sec:z-distort}, the HOD is also effective in modeling the
non-linear and the large-scale bias.

\subsection{The SDSS LRG sample}
\label{sec:sample}

The SDSS Luminous Red Galaxies, as selected by the algorithm developed by
\cite{eisenstein_etal:01}, are an excellent tracer of dark matter on large scales due to 
their high luminosity, which allow us to map them up to $z\sim 0.5$ and comoving volumes up to
1.5 ($h^{-1}$Gpc)$^3$. The selection algorithm, and
the high success rate of the spectroscopy makes (average sky completeness of 98\%) 
 the LRG sample nearly volume-limited up to 
$z\sim 0.35$ and flux limited up to $z\sim 0.47$. 

We use galaxies drawn from a sample comprising 105,831 spectroscopically 
selected LRGs,
based on SDSS Data Release 7 (DR7, \citealt{abazajian_etal:09}). The LRG galaxy selection algorithm is
described in \cite{eisenstein_etal:01}, and the DR7 LRG samples we use
are drawn from the ``DR7-Full'' catalog described in \cite{kazin_etal:09}, which
is publicly available\footnote{http://cosmo.nyu.edu/$\sim$eak306/SDSS-LRG.html}.
 The  DR7-Full sample spans  the redshift range
$0.16<z<0.47$ and includes a range of luminosities in SDSS $g-$band 
of $-23.2<M_g<-21.2$, where the  magnitudes have been K-corrected 
and passively evolved to $z=0.3$, near the median redshift of the sample.
The sample is nearly
volume-limited for redshifts $z\lesssim 0.36$ and 
drops off due to the flux limit at higher redshift. 
The sample covers a sky area of 7908 deg$^2$ (close to 20\% of the sky)
 and  includes data from
both the northern and southern Galactic hemispheres.

From this main catalog, two important samples are extracted: one that includes all LRGs, 
but is volume-limited, which we call ``DR7-Dim", which has a maximum
redshift of $z=0.36$ and median spatial density of $\sim 10^{-4}$ $(\hmpc)^{-3}$, 
and a sample of the bright LRGs, where $M_g<-21.8$, which
we call ``DR7-Bright", which reaches redshifts up to $z=0.44$ and has 
a low spatial density  $\sim 2.5\times 10^{-5}$ $(\hmpc)^{-3}$.
Both of these samples are the same as the ones presented in
\cite{kazin_etal:09}, and use data from the northern cap only, in order to have more
mock catalogs when estimating the errors and covariance of our measurements (see
\S \ref{sec:sims} for more details).

To calculate the correlation functions, we
use the same DR7-Full random catalogs as in \citealt{kazin_etal:09}, which contain 
10 times the number of LRGs as the data. These random catalogs have been
built having the same redshift selection function as the data, as well as the same
sky completeness. For the purposes of calculating the 3PCF in a reasonable time, on the largest
scales we lower the number of random points; when measuring the 3PCF at $s=30$ $\hMpc$, our largest
scale, our random catalog has 4 times the number of data points.

For comparison, we have also analyzed the earlier SDSS DR3 LRG sample
used by \citealt{kulkarni_etal:07}. That sample covers 3816 deg$^2$
and  contains 50,967 LRGs spanning the redshift range $0.15<z<0.55$. 
In this case, we use the data and random catalogs of 
\cite{eisenstein_etal:05}, kindly provided by the author.

The information on these samples is summarized in Table \ref{tab:samples}.

\begin{figure}
\plotone{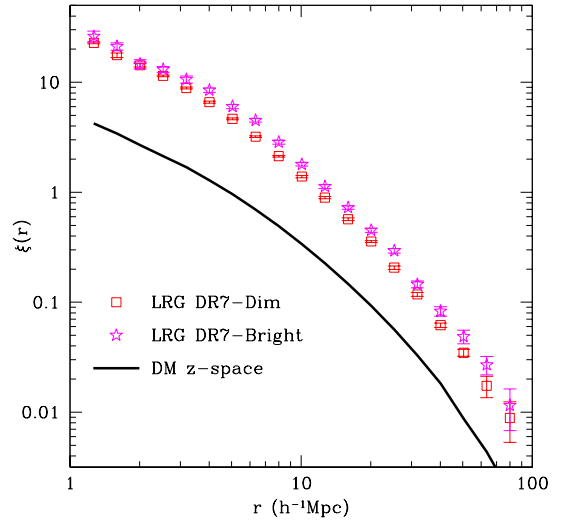}
\caption{The redshift space 2PCF for LRGs. Red open squares correspond to measurements in the DR7-Dim  LRG sample, magenta stars
 represent the 2PCF of the DR7-Bright LRG sample. Thick black solid line corresponds to the dark matter 2PCF, from N-body simulations by
 \cite{sabiu_etal:10}. Errors from the variance of LasDamas LRG mock catalogs.}
\label{fig:lrg2pt}
\end{figure}

The LRGs, i.e., giant red ellipticals, are associated with masssive halos, and they are are highly clustered
\citep[see for instance][]{zehavi_etal:05lrg,masjedi_etal:06}.
This is illustrated in Figure \ref{fig:lrg2pt}, where we show the redshift space 2PCF
of the DR7-Dim and DR7-Bright LRGs, along with the dark matter redshift 2PCF from N-body simulations with 
a cosmological parameters based on the WMAP5 results \citep[][described
in \S \ref{sec:sims}]{komatsu_etal:09}. Both LRG samples have a higher 2PCF than the dark matter, and 
on large scales the offset seems constant, and due to this fact we can say that the LRGs are a biased tracer of dark matter
on large scales. The DR7-Bright LRGs  have a higher 2PCF than the DR7-Dim LRGs, suggesting that they are associated
with more massive halos \citep{zheng_etal:08}.

\begin{figure*}
\plotone{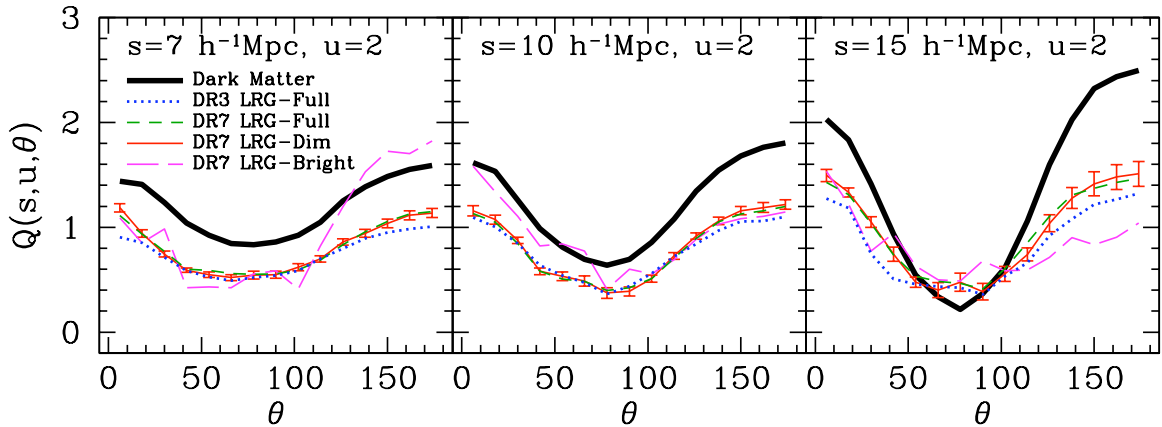}
\caption{The redshift-space, reduced LRG 3PCF,  $Q(\theta)$, for
  configurations with $u=2$ and $s=7$ (left), 10 (middle), and 15
  (right) $\hmpc$. 
\emph{Red solid line:} DR7-Dim LRG sample, errors from  variance of LasDamas mock catalogs.
\emph{Green short-dashed line:} DR7-Full LRG sample. 
\emph{Magenta long-dashed line:} DR7-Bright LRG sample.
\emph{Blue dotted line:} DR3-Full LRG sample.
\emph{Thick black line:} Dark Matter 3PCF from simulations. Error bars from the variance of LasDamas LRG mock catalogs.}
\label{fig:mainLRG}
\end{figure*}

\subsection{Simulations \& Mock catalogs}
\label{sec:sims}

To help analyze and interpret the 3PCF results, 
we use mock catalogs drawn from N-body simulations. The mock 
catalogs serve two functions: (i) to measure the 3PCF of dark matter, 
with which the galaxy 3PCF measurements can be compared to infer the 
bias, 
and (ii) to determine the 
expected error covariance matrix for the 3PCF from the variance among 
a large number of mock catalogs. We use different simulations tailored 
to these different functions.

First, 
to study the dark matter two- and three-point correlation functions, we use six
N-body simulations  carried out by \cite{sabiu_etal:10}, kindly provided by the 
author,  using the GADGET code.
The cosmological model is 
a spatially flat $\Lambda$CDM universe,
with matter density parameter $\Omega_m = 0.27$,  baryon density $\Omega_b=0.045$,
power spectrum amplitude $\sigma_8=0.8$, reduced Hubble parameter $h=0.7$, and initial
power spectrum index $n_s =0.95$. The simulation box volume is 1
($h^{-1}$Gpc)$^3$ and contains $512^3$ dark matter particles 
with a particle mass of $5 \times 10^{11} \hMsun$, with a spatial softening of
60 $h^{-1}$kpc, and we use the output
at $z=0.3$, the mean redshift of the LRGs, to compare their correlation functions.

Since using all the dark matter particles for the 3PCF calculations would take
a prohibitively long time, we randomly select particles from the 
simulation, to a degree where we are sure that the dilution will not
affect the results. For the 3PCF measurement, 
this depends on scale, and we use between 5\% and 0.5\% of the
particles (a higher fraction for 
smaller scales). To obtain the spatial coordinates in the redshift-space dark-matter catalog, we 
use the long-distance approximation in the $z$-direction \citep[see e.g][]{bernardeau_etal:02}
 to account for the effect of peculiar velocities
on the measurement of the correlation functions. 

We use the dark matter halo catalogs obtained from 
these simulations to study  redshift distortions in the estimation of the bias parameters (section \S \ref{sec:z-distort} below), 
found using a Friends-of-Friends algorithm \citep{davis_etal:85} with a linking length of $b=0.2$ times
 the mean interparticle separation.
To populate  halos with mock LRGs, we use
 the best-fit HOD parameters from fits to the projected 2PCF by  
 \cite[][]{zheng_etal:08}
  in their 5-parameter model.
We assign mock galaxy positions, based on halo profiles, and peculiar velocities (based on a gaussian distribution, which depends on the 
mass of the host halo) of galaxies inside halos  using a code kindly provided by Jeremy Tinker.

Second, to estimate the errors in the 3PCF estimates, 
we use the covariance among public mock catalogs from the LasDamas simulations 
\citep{mcbride_etal:10}\footnote{http://lss.phy.vanderbilt.edu/lasdamas/}. 
As will be discussed in \S \ref{sec:errors}, 
we prefer this error estimation method to the use of jack-knife (JK) resampling. 
For the LasDamas simulations, the cosmological model is similar but not
identical to that above: flat, $\Lambda$CDM with $\Omega_m = 0.25$,  
$\Omega_b=0.04$,
$\sigma_8=0.8$, $h=0.7$, and 
$n_s =1$. For this set of parameters, the LasDamas team has carried out N-body simulations
in boxes of different sizes to model galaxies of different luminosities; for the
LRGs, they use  40 boxes with  volume of $2400^3$
$(\hmpc)^3$, each one containing $1280^3$ dark matter particles 
with a particle mass of $4.57\times 10^{11} \hMsun$.

These simulations
are used to construct mock galaxy 
catalogs by placing galaxies in dark matter halos using
a Halo Occupation Distribution (HOD) model \citep{berlind_weinberg:02}, with HOD parameters fit
from the observed SDSS LRG  2-point clustering. From this main set they create two types of galaxy 
mock catalogs, one that resembles the LRG DR7-Dim catalog, with $z_{max}=0.36$, and one
set that resembles the Bright LRG sample with $z_{max}=0.44$. 
 The mocks reproduce the SDSS DR7 geometry, 
and they include redshift-space distortions from peculiar velocities. 
These galaxy mocks have been shown
to reproduce well the form of the DR7 LRG 2PCF on large scales
\citep{kazin_etal:09}, 
and we show below that they are in general agreement with the 3PCF
measurements as well. Since our DR7-Dim and DR7-Bright samples use only the northern cap of the 
SDSS, in each LasDamas box it is possible to extract four mock LRG catalogs;
in this way, we have 160 mock catalogs for each sample.

\section{Results}
\label{sec:lrg3pcf}

 \begin{figure*}
\plotone{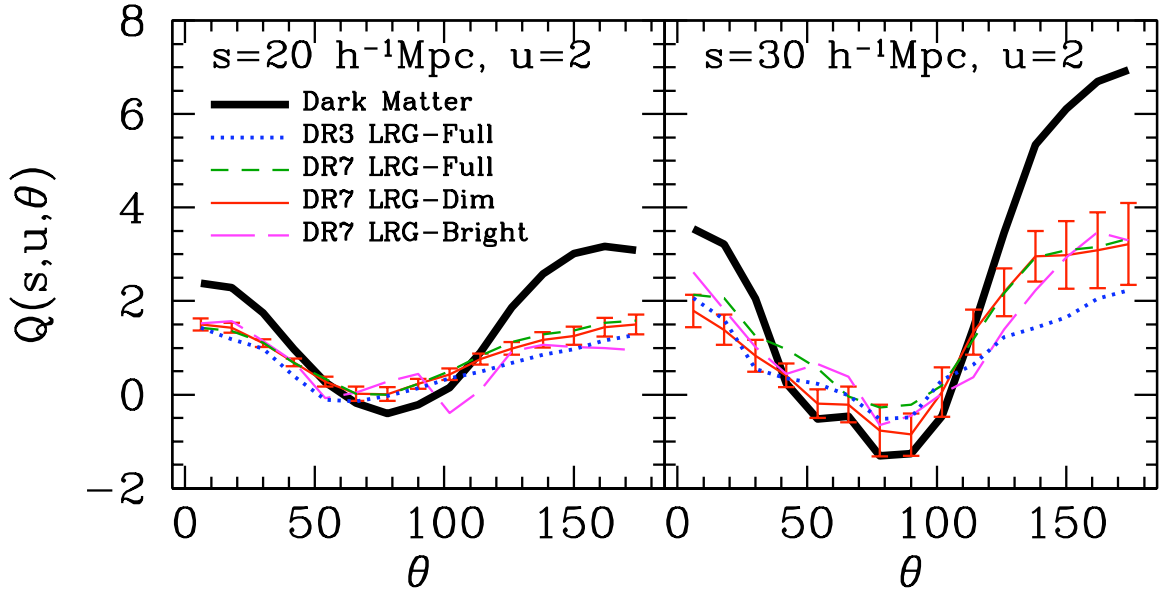}
\caption{The redshift-space, reduced LRG 3PCF,  $Q(\theta)$, for
  configurations with $u=2$ and $s=20$ (left) and 30
  (right) $\hmpc$. 
\emph{Red solid line:} DR7-Dim LRG sample, errors from  variance of LasDamas mock catalogs.
\emph{Green short-dashed line:} DR7-Full LRG sample. 
\emph{Magenta long-dashed line:} DR7-Bright LRG sample.
\emph{Blue dotted line:} DR3-Full LRG sample.
\emph{Thick black line:} Dark Matter 3PCF from simulations. Errors from the variance of LasDamas LRG mock catalogs.
 Note difference in vertical scale compared to 
Figure \ref{fig:mainLRG}.}
\label{fig:bigLRG}
\end{figure*}

We present the redshift-space 
LRG 3PCF measurements in Figures \ref{fig:mainLRG}-\ref{fig:bigLRG}
for configurations with 
$s$ between 7, and 30 $\hmpc$, with $u=r_2/r_1=2$. 
We chose these scales, which go from the mildly non-linear to the
quasi-linear regime, in order to calculate the large-scale bias of LRGs, 
and use only triangles with $u=2$ since that puts all triplet distances (i.e. the sides
of the triangle) above the $s$ selected. If we used $u=1$, then our most collapsed triangles
would have been in the strong non-linear regime, affecting our bias estimates. We could also try triangles
with $u=3$ or more, but then our biggest distances would be comparable to the scale of the
survey, and also  computing  the 3PCF becomes prohibitive
We first discuss the measurements on intermediate and large 
scales and then explore the 3PCF error estimates.

\subsection{LRG 3PCF on intermediate scales}
\label{sec:inter}

Figure \ref{fig:mainLRG} shows the reduced 3PCF
of the  DR7 LRG samples 
on intermediate scales, for $s=7, 10$, and 15  $\hmpc$. For clarity, we show diagonal
error bars only for the DR7-Dim sample, calculating using  the 3PCF variance between the LasDamas mocks.
For comparison, we also show the 3PCF for the DR3 LRG sample.
The solid black curves show the 3PCF of the dark matter from the N-body simulations.

The reduced 3PCF shows the general shape-dependence expected from
gravitational instability theory \citep{bernardeau_etal:02}: the amplitude is
higher for elongated triangle configurations ($\theta \sim 0^\circ, 180^\circ$), 
reflecting anisotropic velocity flows along density gradients. The
configuration dependence is less pronounced on smaller scales, where 
quasi-virialized, more isotropic flows and structures dominate. Since
these measurements are in redshift space, the
configuration dependence also reflects the effects of redshift
distortions due to peculiar velocities.

Since LRGs are biased tracers of the dark matter, the 3PCF of the LRGs
differs from that predicted for the dark matter. On small scales, the
reduced 3PCF amplitude of the LRGs is smaller by a nearly constant
factor  from that of the dark matter. On larger scales, the nature of the
 3PCF bias changes: there is a shape dependence in the offset between 
the LRG and dark matter amplitudes, with the galaxy 3PCF showing 
less configuration dependence than that of the dark matter.

 Comparing the DR7-Full results to the DR3 measurements, which cover half
 the volume of the DR7 sample, we find general consistency in the 3PCF 
amplitude on these intermediate scales. For $s=15$ $\hmpc$, the DR3 
amplitude is slightly lower than that of the DR7 sample for all 
configurations; this is a consequence of the smaller volume sampled to estimate
the correlation functions: as observed in other studies \citep[see][and refs. therein]{marin_etal:08}, 
the 3PCF is more sensitive to the sample volume than the 2PCF.
  
In general, there are no important differences between the DR7-Full and the volume-limited DR7-Dim sample results.
There are differences between these and the DR7-Bright sample: the reduced 
 3PCF for the bright LRGs fluctuates much more and it deviates from the DR7-Dim sample results for the elongated triangles
 (large $\theta$)  for the $s=7$ and 15 $\hmpc$ configurations, and in the collapsed triangles (small $\theta$) for
 the $s=10$ $\hmpc$ configurations.
 Since we have used a binning more tailored to measure the DR7-Dim correlations on the DR7-Bright (a much less dense sample)
 measurements, we expect more fluctuations due to poorer statistics and effects of large structures \citep{nichol_etal:06}; we'll explore
 the variance of bright LRGs in \S \ref{sec:compld} below.

\subsection{LRG 3PCF on large scales}
\label{sec:large}

In Figure \ref{fig:bigLRG} we show the reduced 3PCF for LRGs on large scales, 
for $s=20$  and 30 $\hmpc$, with $u=2.0$. As expected from 2nd-order perturbation
theory, the reduced 3PCF shows a stronger configuration-dependence on 
these scales compared to smaller scales, with a dramatic increase in the difference
in amplitude between elongated and rectangular configurations at $s\geq20$ $\hmpc$.
That behavior is seen for
both the dark matter and the LRG samples. For 
$s=30$  $\hmpc$, the LRG 3PCF dependence in $\theta$ is more asymmetrical between collapsed
and elongated configurations, as it happens with 
the dark matter 3PCF. The errors are greatly increased as well, and there are strong 
anticorrelations, i.e. negative values of $Q(\theta)$, for the rectangular configurations; this is 
again a consequence of the filamentary shape of the large-scale structure.

The large scale 3PCF of the DR7-Bright galaxies is similar to the DR7-Dim 3PCF, with less significant 
fluctuations on larger scales.
On these scales  the finite-volume effects can be observed more clearly in
the 3PCF:
comparing  the DR7-Full and DR7-Dim LRG samples, we can see
that in general they agree within the errorbars, and the small differences are due to a combination
of finite-volume effects  and biasing effects (the DR7-Full sample includes brighter galaxies at large redshifts, 
which  have a different clustering bias than the DR7-Dim sample).
Comparing the  DR3-Full results compared to DR7-Full and even DR7-Dim,
we can see that the DR3-Full 3PCF deviates from the DR7 samples on the elongated
configurations (larger scales).

\subsection{Comparison with LasDamas mocks}
\label{sec:compld}

 \begin{figure*}
\plotone{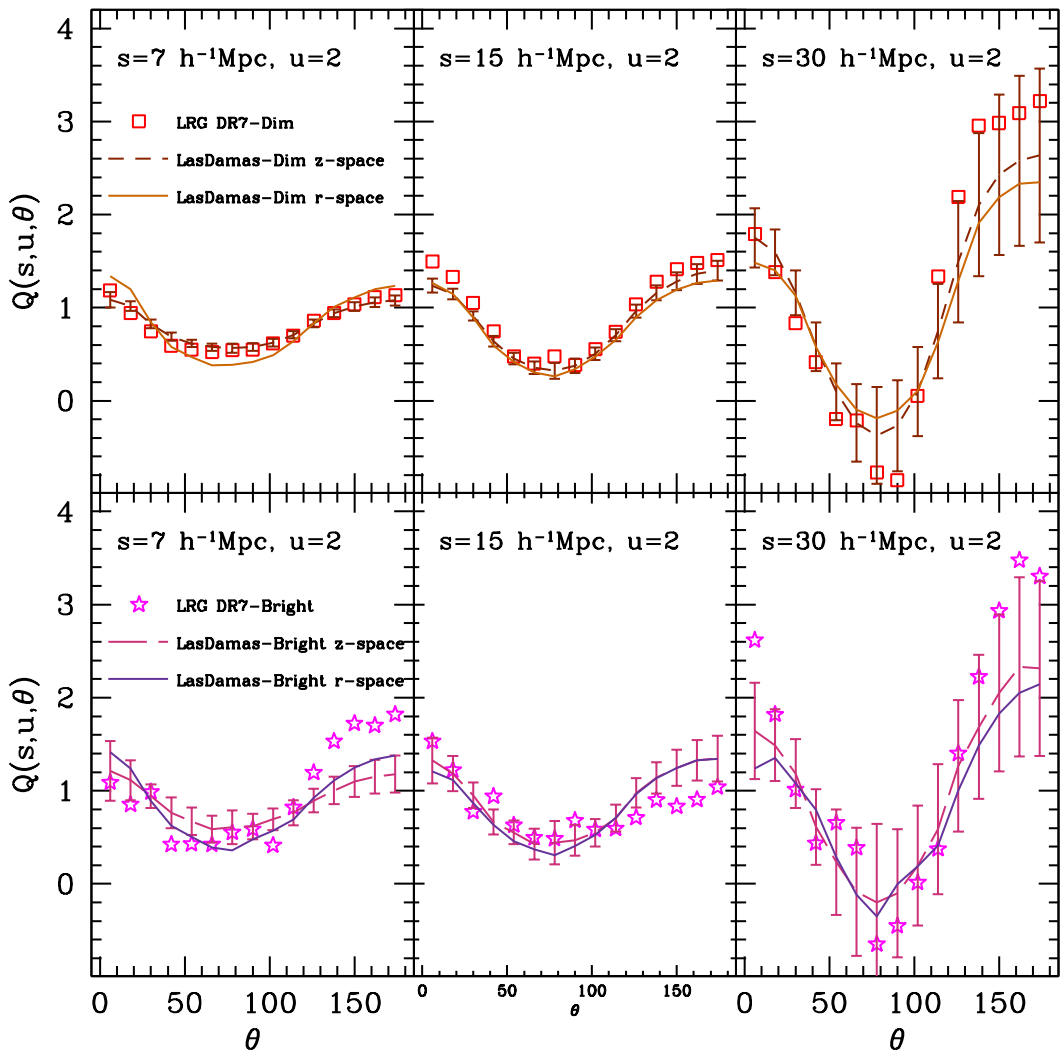}
\caption{Comparison of LRG 3PCF with LasDamas mocks measurements. Symbols represent DR7 measurements, solid lines represent the
mean of the 3PCF measured
in real-space LasDamas catalog, dashed lines represent the mean of the  3PCF measured on redshift-space mocks. 
Error bars represent the variance of the redshift-space 3PCF
measured in 160 mocks.}
\label{fig:comp_mocks}
\end{figure*}

In Figure \ref{fig:comp_mocks} we compare the 3PCF measurements for the DR7 LRGs, as shown previously in
\S \ref{sec:inter} and \S \ref{sec:large}
to the ones obtained from the LasDamas mocks, with dashed lines for redshift-space measurements, solid lines
for real-space measurements.
Here we present a representative set of values, with $s=7$, 15, and 30 $\hmpc$. 
In the top panels we show the comparison for DR7-Dim galaxies; in general the agreement is good,
 and usually within the 1-$\sigma$ error bars.

In the lower panels, we compare DR7-Bright galaxies and the mock 3PCFs. It can be seen that 
the variance of the bright mocks is larger compared to the dim ones, and the agreement between the mocks and the
data is not 
as good, especially on  smaller scales for the elongated configurations. 
Although due to their low  density less signal-to-noise and more fluctuations in $Q(\theta)$ are expected, the variance from 
the mocks should reflect this, and that is not the case for the smaller scales. Since the mocks are built in order to fit the measured
2-point statistics, it can happen that the the mock miss the higher-order statistics, and for future modeling these statistics
have to be considered.

Note also the differences  between the real and redshift space measurements in the LasDamas mocks. Both have a dependence
on shape, but on small scales the real space 3PCF has a more pronounced shape dependence. On bigger scales, especially 
around $s=15$ $\hmpc$, they are very similar within the errors. Comparing these differences to the ones found
by \cite{marin_etal:08}, there are two causes of this: first, the LRGs are galaxies that inhabit massive halos 
 \citep[eg.][among others]{kulkarni_etal:07, zheng_etal:08}, which are in the center of the gravitational potential wells of
 the large matter structures, and therefore they are less affected peculiar velocities. 
Second,
the 3PCF on larger scales is much more affected by the morphology of the structures than from the peculiar velocities, 
and therefore on the largest scales the distortions of the reduced 3PCF will be smaller than on the small scales. There is also
a resolution (or binning) component to this; in very thin bins we might be able to detect better the distortions on large scales.
We will use this similarity between the real and redshift space 3PCF  below  in \S \ref{sec:z-distort} to estimate real-space bias 
parameters from the 3PCF.

\section{3PCF Error estimates}
\label{sec:errors}

Two methods have been commonly employed to estimate errors in
clustering measurements: the variance among mock samples and
jack-knife (JK) sampling. \cite{zehavi_etal:05} showed that
the JK method can be reliable for obtaining the covariance
matrix in the 2PCF, comparing with mock catalogs from independent realizations.
In principle, the errors on the 3PCF 
depend on higher-order correlations up to sixth order \citep[see, e.g.,][]{szapudi:05} 
and could be calculated analytically once the latter are measured. In practice, 
including edge effects and shot noise contributions makes the analytic 
computation of these errors a complicated and computationally challenging problem. 
The jack-knife, in which one sequentially removes subvolumes and
computes the variance among 3PCF measurements for the remaining
volumes, 
provides a convenient computational short-cut, but there is no 
fundamental basis for assuming it is accurate in this context. Indeed, the jack-knife 
method
assumes that the removed subvolumes are independent, ignoring density 
perturbation modes on scales larger than those of the subvolumes. The variance 
in 3PCF estimates among a large number of simulated realizations of a given 
model (mock samples) provides a ground-truth measure of the 3PCF error
for that model. If the model provides an accurate representation of
the actual galaxy clustering, then the mock sample variance should
provide the best estimate of the errors. In the following, we compute
and compare the mock variance estimates of 3PCF errors,
using the 160 LRG-Dim mocks from the LasDamas simulations,  to the 
jack-knife variance estimations from a representative LasDamas LRG-Dim mock. In this
case we divided the mock catalog into many angular regions with equal number of galaxies (at
the limit of large number of subsamples, this is equivalent to have regions with equal volume); each
JK subsample is built by extracting one of these regions from the LRG mock.

\subsection{Diagonal Errors}

For the jack-knife and variance methods, the diagonal errors are given by 
\begin{equation}
\label{eq:diagerr}
\sigma_Q^2 = \frac{E_{method}}{N}\sum_{i=0}^N(Q_i-\bar{Q})^2~.
\end{equation}
For the jack-knife method, $N=N_{JK}$, the number of subvolumes, and 
$E_{JK}=N_{JK}-1$; for the variance among mocks, $N$ is the number of independent 
realizations, and
$E_{var}=1$. 

To have a qualitative estimate of the differences in the diagonal errors between the two methods, we
calculate, for each sample, the average of the reduced 3PCF for all angles in a particular configuration, i.e., we 
calculate $\bar{Q}(s=15,u=2,\theta)$, the mean of $Q(\theta)$ for triangles with $s=15.0$ $\hMpc$, $u=2.0$, as a
function of the number $N$  of samples. We do this for three values of theta, one corresponding to an almost collapsed
triangle ($\theta=18^\circ$), another for a rectangular configuration ($\theta=90^\circ$), and one for an almost
completely extended configuration  ($\theta=160^\circ$). We chose this scale since the 3PCF here is well behaved
and does not have big diagonal errors, but also is large enough to make visible the effects of the JK resampling in the
3PCF (i.e. test the 'independent regions' hypothesis).

\begin{figure*}
\plotone{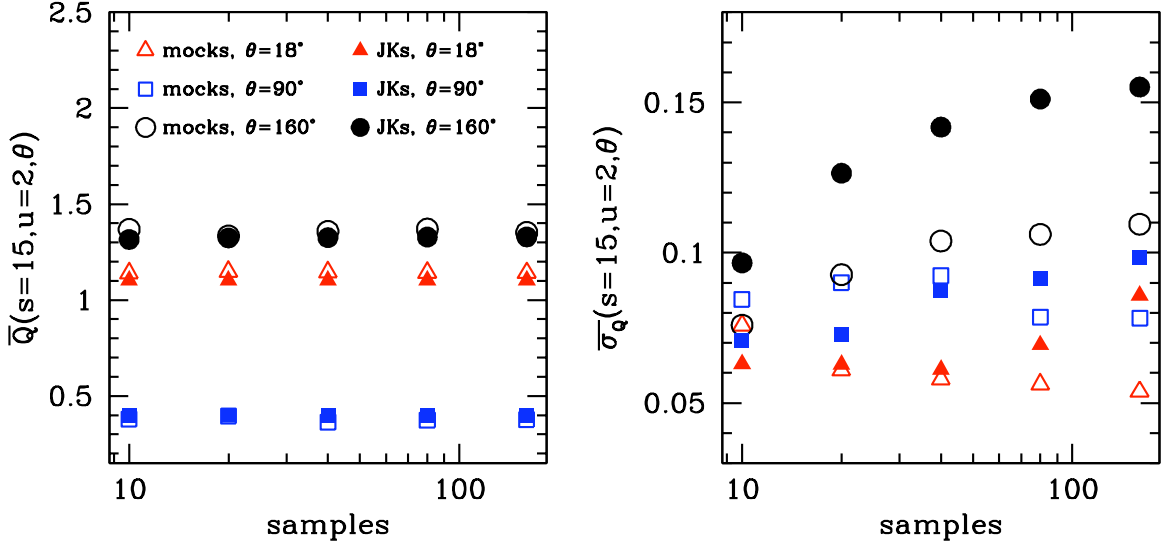}
\caption{Comparison of mean (left) and standard deviation (right) of different reduced 3PCF
 as a function of the number of samples, for
different configurations.
Left:  $\bar{Q}(s=15,u=2,\theta)$, the mean of the $Q(s,u,\theta)$
as a function of number of samples, using LasDamas mocks (open symbols) or 
jack-knife subsamples, 
for triangles with $s=15$ $\hmpc$ and $u=2$, and $\theta=18^\circ$ (triangles), 
$\theta=90^\circ$ (squares), and $\theta=160^\circ$ (circles)
Right: $\bar{\sigma}_Q$, the standard deviation of $Q(\theta)$ as a function of subsamples, from
eq. (\ref{eq:diagerr}).}
\label{fig:comp_s10u2}
\end{figure*}

The results are shown in Figure \ref{fig:comp_s10u2}. In the left panel, we compare $\bar{Q}(s=15,u=2,\theta)$ for
mocks  and for JK subsamples. The average values converge very fast with
$N$, and the fact that the means of the jack-knife samples and the mocks are close 
to each other (differences at the level of 5\% on these scales) is a good signal that we
can use the mocks for the covariance measurements. In the right panel we show the variance (standard deviation) of
$Q(s,u,\theta)$ as a function of $N$, for the different methods. We notice that while the mock's variance
seems to converge at large $N$, the JK errors tend to increase.

\begin{figure*}
\plotone{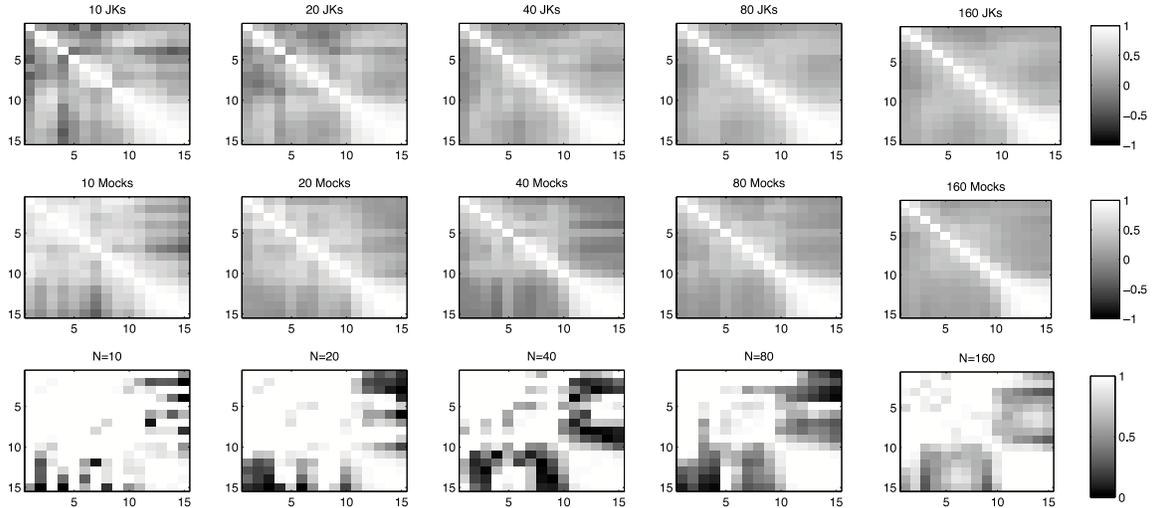}
\caption{
Normalized correlation 
matrices for the reduced, redshift-space 3PCF $Q(s=15 \hmpc, u=2.0,\theta)$ measured in 
LasDamas mock catalogs, using jack-knife resampling for one LasDamas LRG-Dim mock (top),
and the covariance of mocks from independent realizations (middle), for 
$N=10,20,40,80, 160$. The bottom row shows the ratio of each matrix element
$|C^{mocks}_{ij}(N)/C^{JK}_{ij}(N)|$.
 Each matrix element $i,j$
corresponds to the correlation coefficient of 
$Q(\theta=\pi/30 + \pi*(i-1)/15)$ and $Q(\theta=\pi/30 + \pi*(j-1)/15)$}
\label{fig:JKRS_covs}
\end{figure*}

\subsection{Covariance matrices}

For $N$-point correlation functions, measurements for different configurations are strongly 
correlated, therefore the off-diagonal errors must be included in model-fitting and 
parameter estimation. 
Figure \ref{fig:JKRS_covs} shows the correlation matrices of the
3PCF for configurations with $s=15$ $\hmpc$, $u=2$, and varying $\theta$, comparing 
the jack-knife (top row) and mock-covariance methods (middle row),
 for different numbers of jack-knife sub-volumes 
and mock realizations. For both methods, 
the off-diagonal correlation coefficients increase with increasing
$N$, which goes from 10 to 160 subsamples/realizations.

For these configurations, the off-diagonal terms tend to converge when $N\geq40$. As can be seen in the bottom row
of Figure \ref{fig:JKRS_covs}, where we show the ratio between the off-diagonal terms of the covariance matrices,
$|C^{mocks}_{ij}(N)/C^{JK}_{ij}(N)|$,
both
methods yield similar results except for the terms relating very collapsed and very elongated configurations, where the jack-knife
configurations have higher correlations.
This enhanced covariance for the JK methods is due to the fact that this method uses only one 'realization'; while on small 
scales the regions can be effectively independent, on large scales (i.e., when examining large triangles) these regions
are correlated, and then the JK assumption breaks down.  

Even though for large $N$ the covariance matrix converges for the JK method,
and we would be able to extract more information from the covariance matrix,
we would be introducing artificial correlations since the bigger $N$ is, the smaller and ``less independent" the volumes of the subsamples
are.
In our analysis of  large-scale clustering (for bias and cosmological parameters)
we will use the 3PCF error covariance matrices from the variation
among the 160 mock catalogs.

\section{LRG bias}
\label{sec:bias}

\begin{figure*}
\plotone{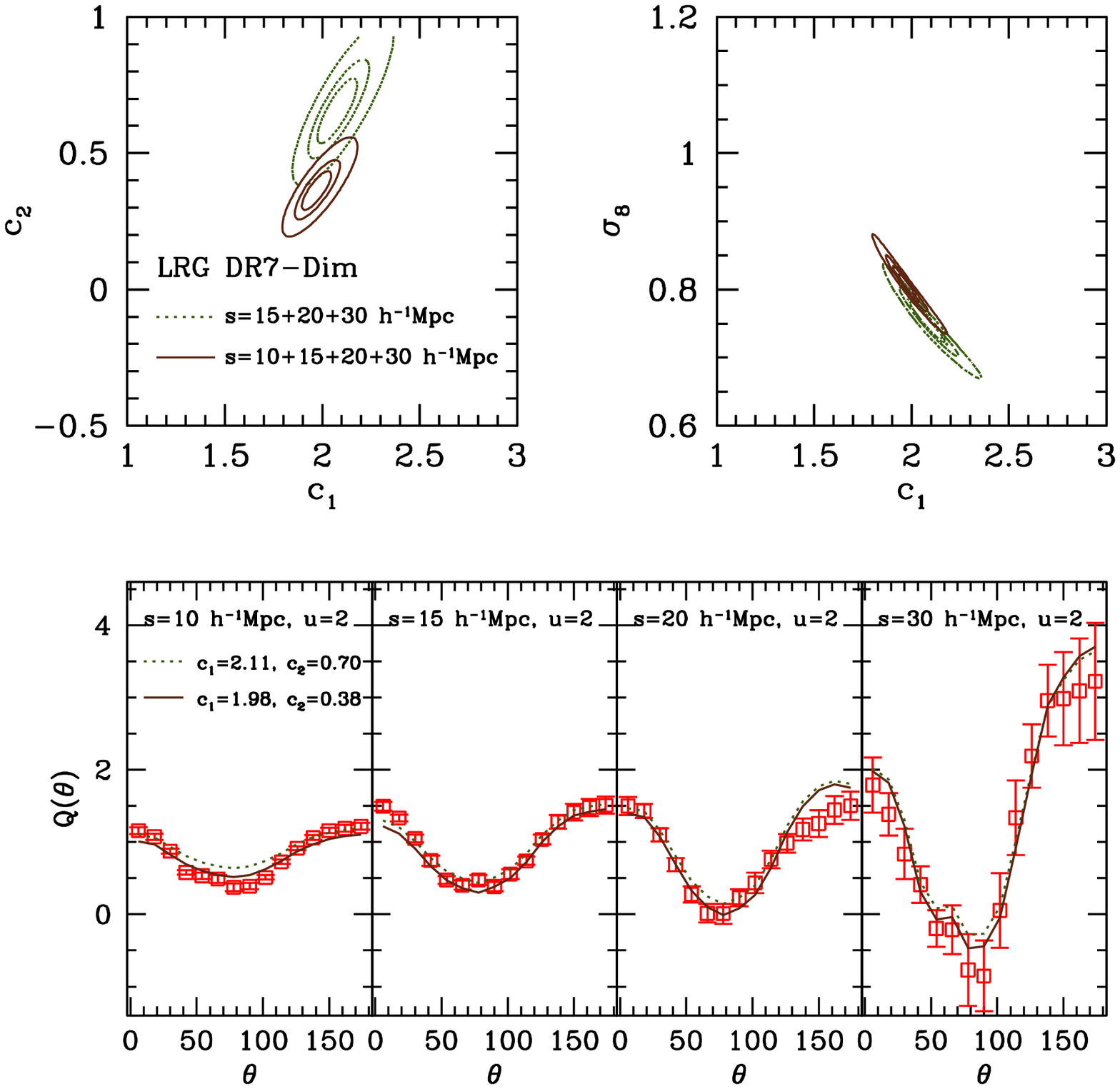}
\caption{Best-fit marginalized likelihood of  bias parameters and $\sigma_8$ from the 2PCF and 3PCF using the LRG DR7-Dim
sample. On the top plots, contours represent
 $\Delta \chi^2=1.0$, 2.3 and 6.2. On the bottom plots, LRG DR7-Dim 3PCF in symbols and lines represent biased dark matter 3PCF values see best-fit
 parameters in Table \ref{tab:biaspar}}
\label{fig:biaspardim}
\end{figure*}

\begin{figure*}
\plotone{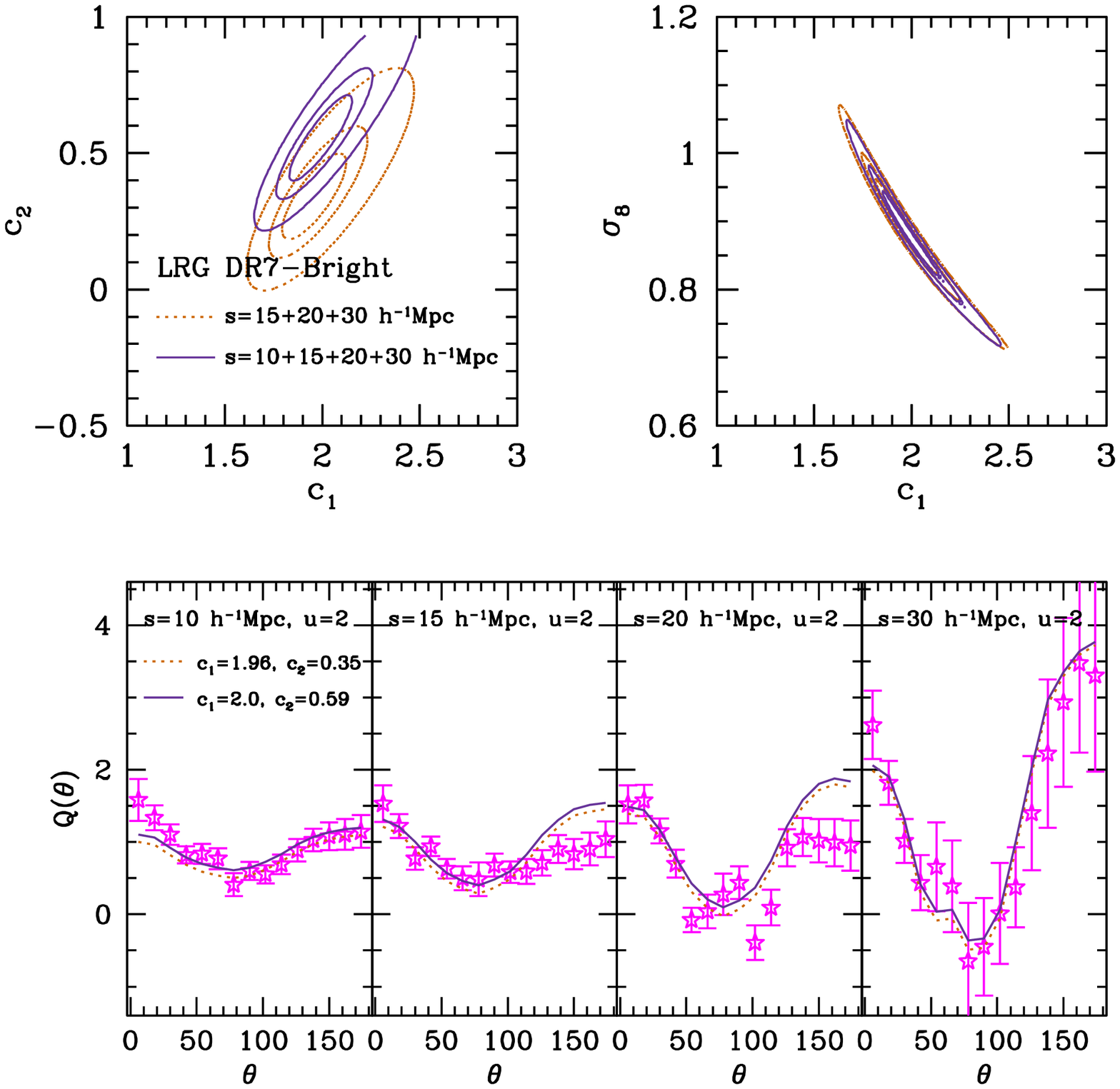}
\caption{Best-fit marginalized likelihood of  bias parameters and $\sigma_8$ from the 2PCF and 3PCF using the LRG DR7-Bright sample. 
On the top plots, contours represent
 $\Delta \chi^2=1.0$, 2.3 and 6.2. On the bottom plots, LRG DR7-Bright 3PCF in symbols and lines represent biased dark matter 3PCF values see best-fit
 parameters in Table \ref{tab:biaspar}}
\label{fig:biasparbright}
\end{figure*}

In the following we quantify the differences between dark matter and LRG clustering,
known as galaxy biasing. As we have seen above, there are important differences 
between them, and, in a similar fashion to what can be done from 2-point 
correlations, we would like to measure the bias of the LRG 3PCF  to obtain cosmological information.
One way to relate dark matter and galaxy clustering in real space is to adopt a 
deterministic and local bias model
\citep[e.g.][]{fry_gaztanaga:93,frieman_gaztanaga:94}, 

\begin{equation}
\delta_{gal}=f(\delta_{dm})= b_1\delta_{dm}+\frac{b_2}{2}\delta_{dm}^2+...,
\label{delta-bias}
\end{equation}
where $\delta_{gal}$ and $\delta_{dm}$ are the local galaxy and 
dark matter overdensities smoothed over some scale $R$. 
To leading order, this bias prescription leads to a relation 
between the galaxy and dark matter reduced 2PCF and connected 3PCF 
amplitudes,
\begin{eqnarray}\label{bias1}
\xi_{gal} (r)&\approx& b_1^2\xi_{dm}(r)\label{eq:2pcfbias}\\
\zeta_{gal}(r_{12}, r_{23}, r_{31}) &\approx&b_1^3\zeta_{dm}(r_{12}, r_{23}, r_{31})+\nonumber \\
 &~& b_2b_1^2[
\xi_{dm}(r_{12})\xi_{dm}(r_{23}) + \textrm{perm.}].\label{eq:3pcfbias}
\end{eqnarray}
In addition to the bias, the 3PCF can help breaking the degeneracy between
the bias and the  r.m.s variance in density of
spheres with radius $R=8$ $\hmpc$,  $\sigma_8$. This parameter is related to the overall amplitude
of the dark matter 2PCF. If we use the 2PCF in equation (\ref{eq:2pcfbias}), adopting 
a fiducial $\sigma_8^{fid}$, then there is a degeneracy between $\sigma_8$ and $b_1$
in linear theory
\citep{pan_szapudi:05}; 
\begin{equation}
\label{eq:bs8}
\xi_{gal}=b_1^2\left(\frac{\sigma_8}{\sigma_8^{fid}}\right)^2\xi_{dm,\sigma_8^{fid}},
\end{equation}
For the 3PCF, there is a factor $(\sigma_8/\sigma_8^{fid})^4$ 
that needs to be multiplied the right side of eq. (\ref{eq:3pcfbias}).
To find the best-fit for the bias parameters and  $\sigma_8$, we use dark matter
correlations in 
a particular fiducial cosmology, and we use
equation (\ref{eq:bs8}) for the 2PCF fit; 
for the reduced 3PCF, in this prescription reads
%
\begin{equation}
\label{eq:biasQ}
Q_{gal} =\frac{1}{c_1}\left(Q_{dm}+c_2\right)
\end{equation}
where $c_1=b_1$ and $c_2=b_2/b_1$, where the dependence on $\sigma_8$ cancels, 
thus allowing us to 
break the degeneracy between $b_1$ and $\sigma_8$ from the 
2PCF. We have to remember, however, that these approximations
and expansions are applicable only on large scales and 
in real space, whereas our measurements are made in redshift space, 
where correlations are distorted due to peculiar velocities. 
Therefore, we have to test the validity of these formulas,
 and try to quantify their effects
on the scales we are using for these comparisons.
We will test how well this 
bias prescription captures the  clustering statistics 
by fitting these relations to the dark matter 
and galaxy 2- and 3-point correlation functions.

\subsection{Constraints on bias parameters and $\sigma_8$ in redshift space}

We  compare the dark matter (from the N-body simulations described in \ref{sec:sims})
and LRG 2PCF and 3PCF in order to constrain
bias parameters using the relations 
(\ref{eq:bs8}) and (\ref{eq:biasQ}), in this case assuming these 
relations are valid in redshift space, i.e. $\xi^{r-space}\rightarrow \xi^{z-space}$
and  $Q^{r-space}\rightarrow Q^{z-space}$.
We use all configurations with $s\geq10$  $\hmpc$, and $s\geq15$  $\hmpc$
 with $u=2$ to measure the joint
likelihood of the bias parameters and $\sigma_8$.

As  described
by \cite{gaztanaga_scoccimarro:05} and used in \cite{gaztanaga_etal:05} 
and \cite{marin_etal:08}, we minimize
\begin{equation}
\chi^2=\sum_{i=1}^{i=2N_b}\sum_{j=1}^{j=2N_b}\Delta_iC_{ij,SVD}^{-1}\Delta_j
\label{eq:chi2}
\end{equation}
where $N_b$ is the number of configurations used ($N_b=60$ for triangles with 
$s\geq 10$ $\hmpc$ and $N_b=45$ for triangles with $s\geq 15 $ $\hmpc$). Since we use both 
2- and 3-point correlations we have
\begin{eqnarray}
\Delta_i &=& (\xi(r_3)^{obs}_i - \xi(r_3)^{model}_i)/\sigma_{\xi(r_3)i}, \textrm{ for } i\leq N_b\\
\Delta_i &=& (Q^{obs}_i-Q^{model}_i)/\sigma_{Q(i)}, \textrm{ for } i>N_b
\end{eqnarray}
where $\xi(r_3)^{model}$ and $Q^{model}$ are given by eqs. (\ref{bias1}) and (\ref{eq:biasQ}) for the galaxies.
The matrix $C_{ij,SVD}$ 
is the normalized
covariance matrix, recalculated with the highest modes from a Singular
Value Decomposition. This covariance matrix is calculated from the LasDamas LRG mock catalogs;
$\sigma_{\xi(r_3)i}$ and
$\sigma_{Q(i)}$ are the uncertainties
in $\xi(r_3)_i$ and $Q(i)$, respectively, and we will assign them the  error bars from the mocks.
We follow the
suggestion of  \cite{gaztanaga_scoccimarro:05} and use 
configurations where the eigenvalues $\lambda < \sqrt{2/N_b}$. As mentioned
above, we actually are more limited by the number of mocks
that we use: we have to use a number of modes smaller than the number
of mocks.

The results are shown in  Figures  \ref {fig:biaspardim} and \ref{fig:biasparbright}, where we use the covariance matrix 
from the 160 LasDamas mocks for each sample to constrain $c_1$, $c_2$, and $\sigma_8$.
In Figure   \ref {fig:biaspardim} we show the results for the DR7-Dim sample, 
and in Figure  \ref{fig:biasparbright}, the results for the
DR7-Bright sample. On the top panels of each figure we show constraints on the bias parameters and
$\sigma_8$, for two sets of configurations: solid lines  use triangles with $s\geq10$ $\hmpc$ 
and dotted lines  use only the largest triangles $s\geq15$ $\hmpc$
The contours correspond to $\Delta\chi^2=1.0$, 2.3 and 6.2 (corresponding to  1$\sigma$ limits for 1-parameter fit,
1$\sigma$ limit for 2 parameter-fit, and 2$\sigma$ limit for 2 parameters, respectively).

We can see that
in both samples a zero non-linear bias is excluded at $2\sigma$ significance.
We can see that, as expected, DR7-Dim galaxies have a lower linear bias $c_1$ than
the DR7-Bright, and there is a bigger difference in the non-linear bias parameter $c_2$. 
Using different
sets of triangles changes the results  affects the area of the confidence
contours, but also the best-fit results; going to smaller scales put us in the direction of the non-linear regime, 
where the relations in eqs. (\ref{eq:bs8}) and (\ref{eq:biasQ}) are not longer valid.
Observe that the error bars are not small: this is an effect of using a large number of
configurations when we have a limited number of realizations (this makes the inverse of the covariance
matrix highly singular, therefore we are obligated to cut many modes), and also
that our errors increase on large scales. 

With respect to the constraints on $\sigma_8$, in general they agree with the WMAP5 results \citep{komatsu_etal:09}, 
but they are more dependent
on which configurations are used. 

The best fit values and 1-$\sigma$ errors are also presented in Table
\ref{tab:biaspar}. 

In the lower two panels we show the LRG 3PCF along with the Dark Matter
biased 3PCF, using the best-fit bias values mentioned above.
For the DR7-Dim sample the fits are quite good, and there
are not significant differences between the fits from using different triangles (solid line 
corresponds to fit using $s\geq10$ $\hmpc$ 
triangles whereas dotted lines use only $s\geq15$ $\hmpc$ triangles). For the brighter sample, the fits are less good
for configurations with $\theta> 100^\circ$ deg.

\begin{figure}
\plotone{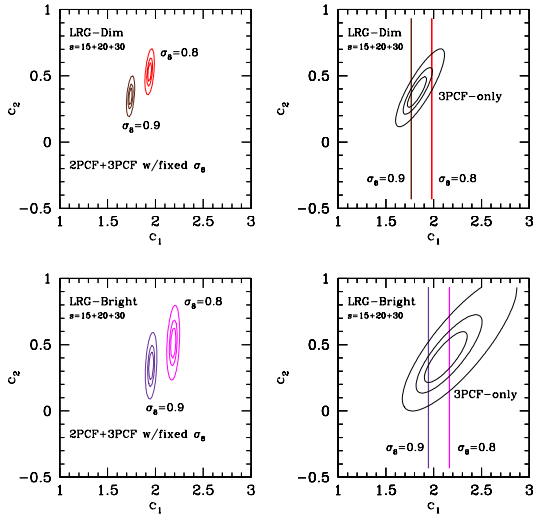}
\caption{Best-fit bias parameter constraints using correlation functions separately and with fixed $\sigma_8$. Left plots
show bias constraints using the 2PCF and 3PCF when fixing $\sigma_8$, right plots show in solid straight lines, the best
fit for $c_1$ using only the 2PCF when $\sigma_8$ is fixed, and the ellipses show constraints in $c_1$ and $c_2$ using the
3PCF only. Upper panels shows the constraints  for the DR7-Dim sample. Lower panels show constraints for the DR7-Bright sample.
Contour lines represent  $\Delta \chi^2=1.0$, 2.3 and 6.2}
\label{fig:bias_fixs8}
\end{figure}

We explore how the constraints
on the bias parameters change if $\sigma_8$ is fixed (instead of being marginalized),
 and compare results to
the analytical bias prediction from the HOD models.
In Figure \ref{fig:bias_fixs8} we show these results, using
only $s\geq15$ $\hmpc$ configurations. On the left side, the vertical ellipses correspond
to contraints on the bias parameters using both the 2PCF and 3PCF, using a model
where $\sigma_8$ is fixed.
As expected, a lower $\sigma_8$ means a higher $c_1$. The constraints on
$c_1$ are tight, due to the lower errors in $\xi$, leaving mostly the uncertainty on $c_2=b_2/b_1$. 

We compare these bias parameters with that expected from the analytical predictions from the best-fit
HOD from \cite[][using projected 2PCF only]{zheng_etal:08} for both
linear and nonlinear bias in the last row of Table \ref{tab:biaspar}, and we have good agreement.
Our error bars, in the case of $\sigma_8=0.8$, confirm that a zero non-linear bias parameter is 
rejected at the 2-$\sigma$ level.
On the right side, we show the bias constraints using the 3PCF only. As can be  seen from the results shown in
Table \ref{tab::biaspar}, 
there is a slight disagreement in the best-fit values using only the 3PCF vs. using both the 2PCF and 3PCF.

\begin{deluxetable*}{l l l c c c}
\tablecolumns{6}
\tablecaption{Best-fit bias parameters in redshift space}
\tablehead{
\colhead{Sample}& \colhead{Triangles used} & \colhead{Fit} & \colhead{$c_1$} & \colhead{$c_2$} & \colhead{$\sigma_8$}}
\startdata
DR7-Dim &  $s\geq10$ $\hmpc$ &2PCF \& 3PCF, 3-parameter  & $1.92^{+0.2}_{-0.1}$ & $0.38^{+0.02}_{-0.08}$ & $0.82^{+0.05}_{-0.13}$ \\
DR7-Dim &  $s\geq15$ $\hmpc$ &2PCF \& 3PCF, 3-parameter  & $2.02^{+0.2}_{-0.1}$ & $0.58^{+0.12}_{-0.08}$ & $0.76^{+0.05}_{-0.07}$ \\
DR7-Bright &  $s\geq10$ $\hmpc$ &2PCF \& 3PCF, 3-parameter  & $1.92^{+0.2}_{-0.1}$ & $0.55^{+0.12}_{-0.15}$ & $0.84^{+0.04}_{-0.05}$ \\
DR7-Bright & $s\geq15$ $\hmpc$ & 2PCF \& 3PCF, 3-parameter  & $1.95^{+0.2}_{-0.1}$ & $0.27^{+0.15}_{-0.08}$ & $0.87^{+0.06}_{-0.07}$ \\
DR7-Dim & $s\geq15$ $\hmpc$ &2PCF \& 3PCF, $\sigma_8=0.9$  & $1.72^{+0.02}_{-0.02}$ & $0.31^{+0.05}_{-0.05}$ & -  \\
DR7-Dim & $s\geq15$ $\hmpc$ &2PCF \& 3PCF, $\sigma_8=0.8$ & $1.93^{+0.02}_{-0.02}$ & $0.45^{+0.05}_{-0.05}$ & -  \\
DR7-Dim & $s\geq15$ $\hmpc$ &3PCF 2-parameter  & $1.83^{+0.12}_{-0.1}$ & $0.32^{+0.12}_{-0.05}$ & - \\
DR7-Bright & $s\geq15$ $\hmpc$ &2PCF \& 3PCF, $\sigma_8=0.9$  & $1.95^{+0.02}_{-0.02}$ & $0.32^{+0.1}_{-0.1}$ & -  \\
DR7-Bright & $s\geq15$ $\hmpc$ &2PCF \& 3PCF, $\sigma_8=0.8$ & $2.18^{+0.02}_{-0.02}$ & $0.51^{+0.1}_{-0.1}$ & -  \\
DR7-Bright & $s\geq15$ $\hmpc$ &3PCF 2-parameter  & $2.05^{+0.1}_{-0.1}$ & $0.48^{+0.1}_{-0.1}$ & - \\
DR7-Dim & - & Z08 HOD fit $\sigma_8$=0.8& $2.22$ & $0.55$ & - \\
DR7-Bright & -& Z08 HOD FoF fit $\sigma_8$=0.8& $2.36$ & $0.79$ & - \\
\enddata
\label{tab:biaspar}
\end{deluxetable*}

\subsection{LRG bias in real space}
\label{sec:z-distort}

\begin{figure*}
\plotone{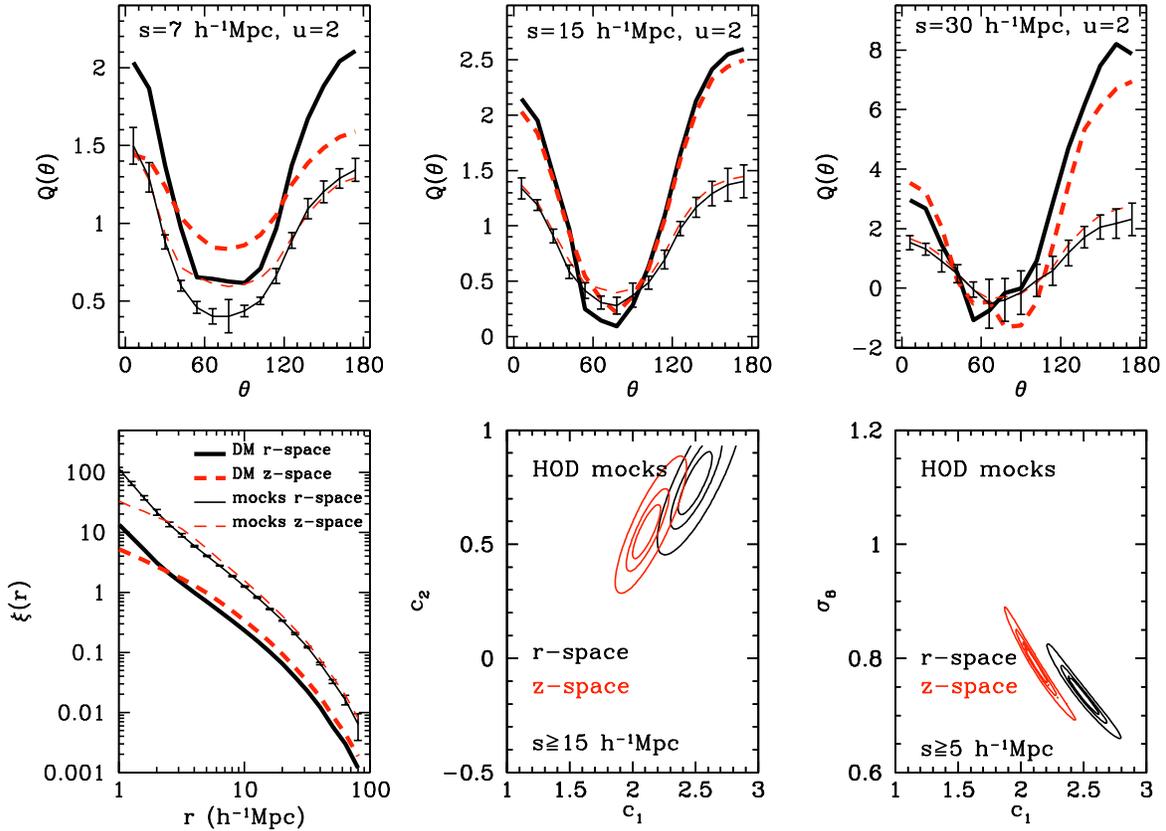}
\caption{Redshift distortions in the 2PCF, 3PCF and in bias constraints. $\emph{Top panels:}$ 3PCF for different
configurations, for dark matter in real (thick black solid line) and redshift space (thick red dashed line), and for  HOD
mocks in real (thin black line, with error bars) and redshift space (think dashed line). 
\emph{Bottom left:} 2PCF for dark matter and mocks in real and redshift space. \emph{Bottom, middle and right:}
Constraints in bias parameters $c_1$ and $c_2$, and in $\sigma_8$ in real and redshift space.
 The contours are  $\Delta \chi^2=1.0$, 2.3 and 6.2}
\label{fig:z_dist}
\end{figure*}

We explore here the validity of using redshift-space correlations to calculate
galaxy bias parameters. 
When mapping the galaxies in the sky, we use the redshift of the galaxy as an indicator of distance. 
But since galaxies (and the dark matter) are subject to gravitational forces from the surrounding structures, 
peculiar velocities are added to the cosmological redshift, distorting our mapping of the large scale structure
and as a consequence, the correlation functions are affected as well \citep[see][for a review]{hamilton:97}.

In the case of the 3PCF, there are differences between the real and redshift space 3PCF of galaxies and dark matter
\citep[eg.][]{gaztanaga_scoccimarro:05, marin_etal:08}, and they will affect bias parameters as well.
\cite{marin_etal:08}, using mock galaxy catalogs,  found good agreement, but not perfect, 
 on the bias parameters in the $L_*$ range, between using real
or redshift space correlations, using $s \sim 10$ $\hmpc$ configurations, 
and  poorer agreement for brighter galaxies. That
was caused by a combination of studying the bias using smaller scales (in the mildly non-linear regime) and
 by the small size of the box (only 200 $\hmpc$ on the side) for studying these galaxies which have a low
 spatial density. 
  
As shown in Figure \ref{fig:comp_mocks}, on large 
scales (with $s\geq15$ $\hmpc$), for the mocks, the difference between the redshift and real space 3PCF
is small and falls within the errors. In the case of the dark matter, there are differences,
 but
because of the large bias, the mocks indicate that the LRG 3PCF will not show significant differences between the redshift and real space
measurements on large scales.
 
In Figure  \ref{fig:z_dist} we show the differences in the correlation functions of dark matter and mock LRG galaxies from 
the N-body simulations by \cite{sabiu_etal:10}, and how
the bias parameters and $\sigma_8$ constraints change from using real and redshift space correlations. 
Note that on small scales
there are significant differences between redshift and real space 3PCFs, but on larger scales, they are very similar.
We have to 
keep in mind that these similarities occur  this binning scheme, 
and might not be necessarily true
with different 3PCF resolutions. For the 2PCF, the differences are small, but significant.
For  the bias constraints, note that real space measurements have a higher linear and non-linear bias, 
and usually the best-fit value
of $\sigma_8$ is lower in the real space fits.

\begin{figure*}
\plotone{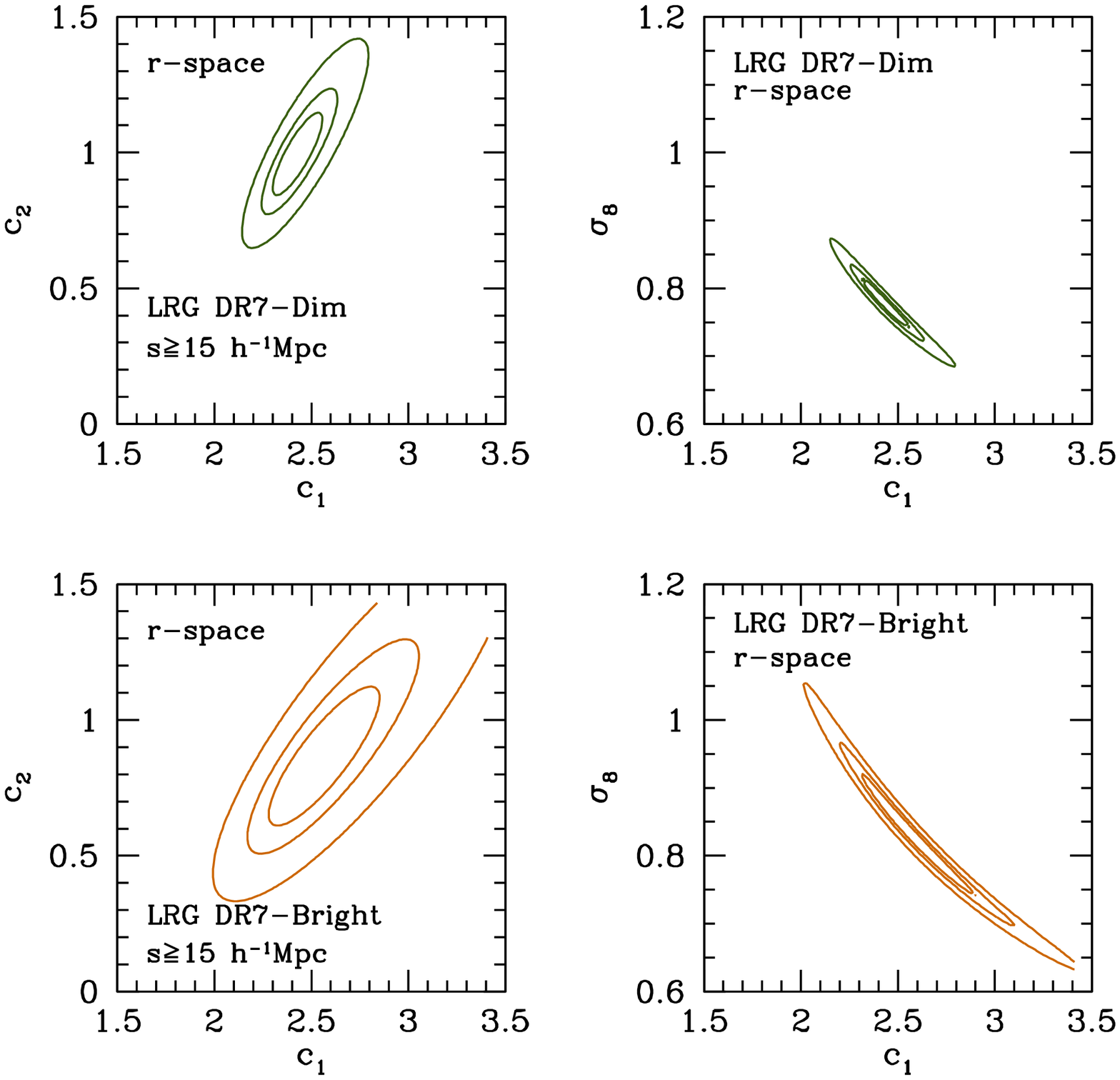}
\caption{Marginalized bias parameters and $\sigma_8$ confidence intervals from joint 2PCF-3PCF fit for LRGs  in real 
space for the DR7-Dim (top) DR7-Bright (bottom) LRG samples using the prescription described in \S \ref{sec:z-distort}. 
Only  $s\geq15$ $\hmpc$ configurations are used.
The contours are  $\Delta \chi^2=1.0$, 2.3, and 6.2.}
\label{fig:lrgbiasR}
\end{figure*}

There have been
analytical attempts to relate the real and redshift space 3-point function, 
in the case of the bispectrum \citep{scoccimarro_etal:99,smith_etal:08}, but there are practically no attempts of this in the 
configuration space, since non-linearities and mixing of large and small-scale modes 
makes the calculations very complicated unless one takes very 
simplified cases or do not take full advantage of the shape dependence of the 3-point statistics \citep[eg.,][]{pan_szapudi:05,sefusatti_etal:06}
In our case, we take an empirical approach.

Using the similarity between the mock LRG 3PCF in real and redshift space for the large scales, we estimate constraints 
on the bias and $\sigma_8$ parameters
using the real space dark matter correlations, and identifying the real space 3PCF with the redshift space 3PCF. In the
case of the 2PCF, since the differences between real and redshift space correlations are significant, we will use the first-order
linear  approximation in the case of redshift distortion \citep{kaiser:87}. Following \cite{pan_szapudi:05}:
\begin{equation}
\xi^{z-space}_{gal}=\left(1+\frac{2}{3}f+\frac{1}{5}f^2\right)c_1^2\left(\frac{\sigma_8}{0.8}\right)^2\xi_{dm}^{r-space},
\label{eq:zdist2pcf}
\end{equation}
where $f=\Omega_m^{0.6}/c_1$; we tested this prescription for our HOD mocks and it works well on the largest scales $r>10$ $\hmpc$.
The results are shown in Figure \ref{fig:lrgbiasR}, for both the LRG-Dim and LRG-Bright samples, using triangles with $s\geq 15$ $\hmpc$.
In general we see the same trend 
that was seen in Figure \ref{fig:z_dist}: both $c_1$ and $c_2$ are higher in real space,
 and we get reasonable constraints on $\sigma_8$.
We notice that here the constraints are in good agreement  with the values obtained from the 
HOD fit of by \cite{zheng_etal:08}  (shown in Table \ref{tab:biaspar}).

\begin{deluxetable*}{l l l c c c}
\tablecolumns{6}
\tablecaption{Best-fit bias parameters in real space}
\tablehead{
\colhead{Sample}& \colhead{Triangles used} & \colhead{Fit} & \colhead{$c_1$} & \colhead{$c_2$} & \colhead{$\sigma_8$}}
\startdata
DR7-Dim &  $s\geq10$ $\hmpc$ &2PCF \& 3PCF, 3-parameter  & $2.51^{+0.15}_{-0.1}$ & $0.82^{+0.12}_{-0.04}$ & $0.75^{+0.03}_{-0.02}$ \\
DR7-Dim &  $s\geq15$ $\hmpc$ &2PCF \& 3PCF, 3-parameter  & $2.38^{+0.2}_{-0.1}$ & $0.92^{+0.1}_{-0.1}$ & $0.77^{+0.02}_{-0.05}$ \\
DR7-Bright &  $s\geq10$ $\hmpc$ &2PCF \& 3PCF, 3-parameter  & $2.22^{+0.2}_{-0.1}$ & $0.75^{+0.2}_{-0.1}$ & $0.91^{+0.1}_{-0.1}$ \\
DR7-Bright & $s\geq15$ $\hmpc$ & 2PCF \& 3PCF, 3-parameter  & $2.45^{+0.45}_{-0.25}$ & $0.79^{+0.4}_{-0.1}$ & $0.85^{+0.1}_{-0.1}$ \\
\enddata
\label{tab:biaspar}
\end{deluxetable*}

\subsection{HOD constraints from large-scale bias}

\begin{figure}
\plotone{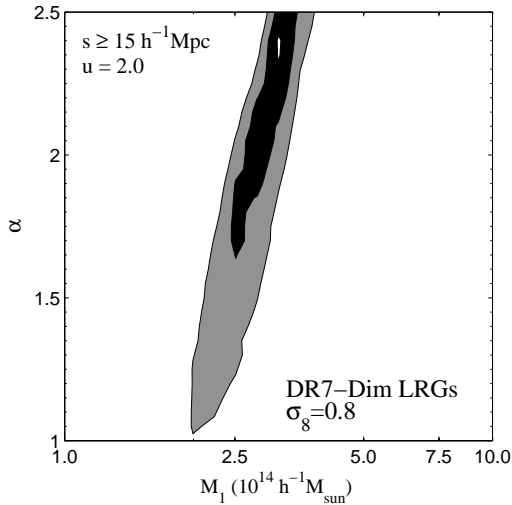}
\caption{Constraints in HOD parameters $M_1$ and $\alpha$ from real-space bias measurements,
 when triangles with
$u=2$ and $s\geq10$ $\hmpc$  are used. The white point is the position of the best-fit HOD parameters.
The contours are  $\Delta \chi^2=2.3$, and 6.2, from inside out}
\label{fig:hodcont}
\end{figure}
It is possible to  use the bias parameters constraints  to estimate best-fit values  on the LRG HOD parameters.
In Appendix \ref{app:biashod} we show how it is possible to extract the large-scale bias parameters from the HOD, where only
the average of the HOD function $\langle N(M) \rangle$ is needed to obtain the bias parameters $b_1^h$ and
$b_2^h$. Given the uncertainties in $c_1$, $c_2$ from the 3PCF, and for a fixed $\sigma_8$, it is possible in principle 
to obtain constraints on
HOD parameters \citep{sefusatti_scoccimarro:05} from large-scale bias parameters. We exemplify this using the LRG DR7-Dim sample.

There are many models for the mean $\langle N(M) \rangle$, with different number of parameters. It is worth mentioning
that even though the meaning of common HOD parameters such as $M_{min}$, $M_1$ and $\alpha$ is similar for
different functional forms of $\langle N(M) \rangle$, the best fit parameters can change significantly depending on the function used
\citep{zheng_etal:05}.
For the
sake of simplicity, and to avoid marginalizing over many parameters, we use a simple model that uses three parameters, 
with a soft function for $N_{cen}(M)$ \citep{sefusatti_etal:06,kulkarni_etal:07}:
\begin{eqnarray}
\langle N(M) \rangle&=&N_{cen}(M)+N_{sat}(M),\\
N_{cen} &=& \exp(-M_{min}/M),\\
N_{sat} &=& \exp(-M_{min}/M)\left(\frac{M}{M_1}\right)^\alpha.
\end{eqnarray}
We effectively constrain the HOD parameters $M_1$ and $\alpha$, since 
we fix the spatial density of model galaxies to the mean density of the
 sample, and in practice, this constrains $M_{min}$. 
 
 We create a grid with different values of $M_1$ and $\alpha$,
 where we calculate the bias parameters  $b_1^h=c_1$ and $b_2^h=c_1c_2$ (see Appendix \ref{app:biashod}), and then we relate
 and use the results for the constraints on the bias parameters $c_1$ and $c_2$ of the DR7-Dim sample when we fix $\sigma_8=0.8$. 
 This gives us a likelihood on the values of $M_1$ and $\alpha$ HOD parameters.
 
In Figure \ref{fig:hodcont} we show the constraints on $M_1$ and
$\alpha$ using the constraints on $c_1$ and $c_2$ for the DR7-Dim LRG sample, using $s\geq15$ $\hmpc$ configurations, 
for fixed $\sigma_8=0.8$.  We  notice the constraints are not very strong, especially on $\alpha$. The  best-fit HOD parameters, 
$M_1=3.2\times10^{14}$ $\hMsun$, and $\alpha=2.35$, and the constraints on $M_1$ 
are consistent with what is expected from fits to 2-point correlations, and although our best fit value for $\alpha$ is high, it is consistent
with the findings of high $\alpha$ for the LRG HOD \citep{blake_etal:08, zheng_etal:08}, and higher from the estimates
from 3PCF fits by \citealt{kulkarni_etal:07}. Note that the strongest
 constraints are on $M_1$, and comparing the shape of these contours with the contours of Figure \ref{fig:bias_fixs8},
we can conclude that $M_1$, the ``average" mass of a halo containing at least two LRGs (a central and a satellite) is mostly
controlled by the linear bias $c_1$, and $\alpha$ is more sensitive to the $c_2$  constraints.

\section{Conclusions and Future Directions}
\label{sec:conclusions}

The purpose of this work was to show that the LRG 3PCF provides  a complementary way to estimate biasing and place constraints
on cosmological models. With the advent of new and larger
surveys,  it is possible to get constraints comparable to other
methods, in cosmology and galaxy population models. Our main findings are:

\begin{itemize}
\item With a new binning scheme, it is possible to have a better measurement of $Q(\theta)$ and
go to larger scales, making it possible to use the LRG 3PCF to learn about
galaxy bias and $\sigma_8$.

\item The measurement of error in the 3PCF is non-trivial. From measuring the 3PCF both in mocks and
JK subsamples, we find that the covariance matrices have noticeable differences, and
JKs tends to overestimate the errors and covariance compared to using mocks.

\item  Using the redshift space 2PCF and 3PCF, we estimate constraints on the linear and 
non-linear bias parameters. We show that the non-linear bias is non-zero at the $2\sigma$ level for LRGs.

\item Using N-body simulations, for this  binning scheme, we find that the predicted LRG 3PCF is very similar (within the
errors) for real and redshift space measurements on large scales.
 Using this fact, we estimate bias parameters in real space, as well
as $\sigma_8$, which gives similar results to WMAP5 estimates.

\item From the large-scale bias parameter constraints, 
we estimate HOD parameters. Our constraints are not
very strong, but the best-fit values are similar to what is found  by other methods.

\end{itemize}

Using a new binning scheme that preserves the shape dependence of $Q(\theta)$, it is  possible to
measure the 3PCF on a larger range of scales with a better signal-to-noise than previous schemes.
We treated here the statistical errors only, but future investigations (outside the scope of this particular work) need
also to address the issue of how different resolutions, or binning schemes, can lead to systematic errors in the 3PCF
measurements and on bias constraints.

The measurement of errors shows that, provided that the mocks we use describe adequately the correlations
of the sample which is analyzed, it is much preferable to use these instead of JK resampling as a method to calculate the
covariance of  3PCF measurements. As opposed  to the case of the 2PCF \citep[eg][]{zehavi_etal:05}, 
where the JK provide a good description, in the 3PCF we would need much larger volumes and even there, 
measure only on a limited range of
scales, to be able to trust the JK error measurements. This is more of a problem when we are talking about the covariance
of the 3PCF measurements. For diagonal errors, it is possible to obtain them using a small number of subsamples 
$N\sim10$, but to have a converged covariance matrix,  $N\gsim100$ are needed.

The aspects mentioned above will affect the model constraints from the 3PCF, and we will work
on these aspects as well in future investigations. We showed that LRGs, or equivalently, high-mass
galaxies have a high (non-zero) positive non-linear bias, and a confirmation from what is predicted by the halo and HOD models:
that high mass galaxies (and halos) have large values of the linear and nonlinear bias parameters.

For larger scales, we  take advantage of the fact that the LRG real space 3PCF is similar to the redshift space counterpart. 
With better measurements this might not be possible to do, especially if the purpose is to obtain more
precise constraints on $\sigma_8$, but the constraints obtained are consistent with what is expected from simulations, and 
moreover, they allow us to put complementary constraints on galaxy occupation models. 

The future of the LRG 3PCF looks promising: apart from better constraints on biasing, on small scales HOD fits will become
relevant, and on the largest scales, constraints from the BAO will also contribute to constraining cosmological and
galaxy formation models.

\acknowledgements

The author thanks Joshua Frieman, for his encouragement in the research of  high-order 
clustering in both theory and observations. Also many thanks to thank Cristiano Sabiu for 
facilitating his dark matter halo catalogs, to Jeremy Tinker for making available his HOD code, and to
Daniel Eisenstein and Eyal Kazin for making their LRG data and random catalogs available, and to 
Roman Scoccimarro for making his dark matter halo catalogs available in the early steps of this research.
The author is grateful to Robert Nichol  and Risa Wechsler, for their  initial and continuing
 inspiration to study the LRG 3PCF.
Additional thanks to thank Hsiao-Wen Chen, Scott Dodelson, and Andrey Kravtsov, Enrique 
Gazta\~naga, Emiliano Sefusatti, Idit Zehavi, Chris Blake and Cameron McBride for useful suggestions and comments.
This work was supported in part by the Kavli Institute for Cosmological Physics 
at the University of Chicago through grants NSF PHY-0114422 and NSF PHY-0551142 
and an endowment from the Kavli Foundation and its founder Fred Kavli. 
Some of the calculations used in this work have been performed on the Joint Fermilab-KICP Supercomputing Cluster, supported by grants from Fermilab, Kavli Institute for Cosmological Physics, and the University of Chicago. The author thanks
ICG Portsmouth and KIPAC/SLAC for their hospitality where initial stages of this work took place.

 Funding for the SDSS and SDSS-II has been provided by the Alfred P. Sloan Foundation, the 
 Participating Institutions, the National Science Foundation, the U.S. Department of Energy, the National 
 Aeronautics and Space Administration, the Japanese Monbukagakusho, the Max Planck Society, and 
 the Higher Education Funding Council for England. The SDSS Web Site is http://www.sdss.org/.

   The SDSS is managed by the Astrophysical Research Consortium for the Participating Institutions. The Participating Institutions are the American Museum of Natural History, Astrophysical Institute Potsdam, University of Basel, University of Cambridge, Case Western Reserve University, University of Chicago, Drexel University, Fermilab, the Institute for Advanced Study, the Japan Participation Group, Johns Hopkins University, the Joint Institute for Nuclear Astrophysics, the Kavli Institute for Particle Astrophysics and Cosmology, the Korean Scientist Group, the Chinese Academy of Sciences (LAMOST), Los Alamos National Laboratory, the Max-Planck-Institute for Astronomy (MPIA), the Max-Planck-Institute for Astrophysics (MPA), New Mexico State University, Ohio State University, University of Pittsburgh, University of Portsmouth, Princeton University, the United States Naval Observatory, and the University of Washington.

This work is presented as a dissertation to the Department of Astronomy and Astrophysics, The University of Chicago,  in partial fulfillment of the requirements for the Ph.D. degree.

\bibliography{arxivms}

\begin{appendix}
\section{LRG HOD and large-scale bias}
\label{app:biashod}

\subsection{Halo Mass Function}

The distribution of dark matter  halos is biased with respect to the dark matter \citep{press_shechter:74}, 
therefore the galaxies, formed inside these halos,  are biased tracer as well. As mentioned above, in 
the HOD model we proceed in two steps: First, we described how dark matter halos are distributed in the 
universe, and second, how galaxies populate these halos.

To describe the mass function of dark matter halos (i.e., the spatial density of halos as a function of mass) we 
use the Sheth \& Tormen mass function \citep{sheth_etal:01, sheth_tormen:02}:
\begin{equation}
n_h(M) = \frac{\bar{\rho}_m}{M^2}\left|\frac{d\ln\sigma}{d\ln M}\right| f(\nu)
\end{equation}
where $\bar{\rho}_m\equiv\Omega_M\rho_c$  is the average matter density in the universe at the epoch of observation, 
$\nu=\delta_c/\sigma(M,z)$, with $\delta_c=1.69$ the threshold for growth
of linear fluctuations, and  $\sigma$, the r.m.s. variance of a sphere of radius $R(M)$ at redshift $z$:
\begin{equation}
\sigma^2(R,z)= \int \frac{d^3k}{(2\pi)^3} \left| W(k,R) \right|^2P(k,z).
\end{equation}
Here $P(k,z)$ is the linear matter power spectrum and $W(k,R)$ is the 
Fourier transform of a top-hat window of radius R. The function $f(\nu)$ is motivated 
by the ellipsoidal collapse model  \citep{sheth_tormen:02} and has the form
\begin{equation}
f(\nu) = A\sqrt{\frac{2a\nu^2}{\pi}}\left[1+(a\nu^2)^{-p}\right]e^{-a\nu^2/2},
\end{equation}
where for normalization $A=0.322$, $p=0.3$ amd $a=0.707$.

\subsection{HOD bias}

Knowing the probability function $P(N|M)$ of how galaxies populate halos as a function of mass, we can infer
the clustering at any scale.
The HOD model allow us, knowing the mean occupation number $\left<N(M)\right>$, 
in the large-scale limit to estimate  the bias parameters:
\begin{equation}
\label{eq:biasint}
b_i = \frac{1}{n_g}\int_{M_{min}}^{\infty} dM n_h(M)b_i^h(M)\left<N_g(M)\right>,
\end{equation}
where i=1,2 represent the linear and first non-linear bias term, 
$n_h(M)$ the dark matter halo mass function, $b_i^h(M)$ is the halo bias and
$\left<N_g(M)\right>$ is the mean number of galaxies for a halo of mass $M$, and
$n_g$ is the galaxy number density, calculated as:
\begin{equation}
n_g = \int_{M_{min}}^{\infty} dM n_h(M)\left<N(M)\right>,
\end{equation}

The dark matter halo bias parameters are given by \cite{scoccimarro_etal:01}
\begin{eqnarray}
b_1^h(M) &=& 1+\epsilon_1+E_1\\
b_2^h(M) &=& \frac{8}{21}\left(\epsilon_1+E_1\right)+\epsilon_2+E_2
\end{eqnarray}
the coefficients given by
\begin{eqnarray}
\epsilon_1 &=& \frac{a\nu^2-1}{\delta_c},\\
\epsilon_2 &=& \frac{a\nu^2(a\nu^2-3)}{\delta_c^2},\\
E_1 &=& \frac{2p}{\delta_c(1+(a\nu^2)^p)},\\
\frac{E_2}{E_1} &=& \frac{1+2p}{\delta_c}+2\epsilon_1.
\end{eqnarray}
It is worth emphasizing that the bias relations mentioned in equation (\ref{eq:biasint}) are only valid on large scales, in the quasi-linear regime, and
they are scale independent. On  smaller scales the bias becomes very scale-dependent and higher moments
of the $P(N|M)$ function need to be known.

\end{appendix}

\end{document}